\documentclass[twocolumn]{aastex63}
\usepackage{graphicx}
\usepackage{color}
\usepackage{amsmath}
\usepackage{mathrsfs}
\bibpunct{(}{)}{;}{a}{}{,}

\received{}
\revised{}
\accepted{September 2, 2021}
\shorttitle{Cosmic Ray-Induced Desorption of Interstellar Ices}
\shortauthors{Sipil\"a et al.}

\graphicspath{{./}{figures/}}

\begin{document}

\title{A Revised Description of the Cosmic Ray-Induced Desorption of Interstellar Ices}

\correspondingauthor{Olli Sipil\"a}
\email{osipila@mpe.mpg.de}

\author[0000-0002-9148-1625]{Olli Sipil\"a}
\affiliation{Max-Planck-Institut f\"ur extraterrestrische Physik, Giessenbachstrasse 1, 85748 Garching, Germany}

\author[0000-0003-1572-0505]{Kedron Silsbee}
\affiliation{Max-Planck-Institut f\"ur extraterrestrische Physik, Giessenbachstrasse 1, 85748 Garching, Germany}

\author[0000-0003-1481-7911]{Paola Caselli}
\affiliation{Max-Planck-Institut f\"ur extraterrestrische Physik, Giessenbachstrasse 1, 85748 Garching, Germany}

\begin{abstract}
Non-thermal desorption of ices on interstellar grains is required to explain observations of molecules that are not synthesized efficiently in the gas phase in cold dense clouds. Perhaps the most important non-thermal desorption mechanism is one induced by cosmic rays (CRs), which, when passing through a grain, heat it transiently to a high temperature -- the grain cools back to its original equilibrium temperature via the (partial) sublimation of the ice. Current cosmic-ray-induced desorption (CRD) models assume a fixed grain cooling time. In this work we present a revised description of CRD in which the desorption efficiency depends dynamically on the ice content. We apply the revised desorption scheme to two-phase and three-phase chemical models in physical conditions corresponding to starless and prestellar cores, and to molecular cloud envelopes. We find that inside starless and prestellar cores, introducing dynamic CRD can decrease gas-phase abundances by up to an order of magnitude in two-phase chemical models. In three-phase chemical models our model produces very similar results to the static cooling scheme -- when only one monolayer of ice is considered active. Ice abundances are generally insensitive to variations in the grain cooling time. Further improved CRD models need to take into account additional effects in the transient heating of the grains, introduced for example by the adoption of a spectrum of CR energies.
\end{abstract}


\section{Introduction}

Observations have revealed a gas-phase presence of complex organic molecules, which have been thought to be synthesized on the surfaces of interstellar dust grains, in cold and dense clouds \citep[e.g.,][]{Cernicharo12,Vastel14,Jimenez-Serra16}. To explain these observations, new gas-phase production routes have recently been explored \citep{Vasyunin13b,Balucani15,Vasyunin17}, but even these depend on the desorption of precursor species from the grain surfaces. Thermal desorption of most species is negligible because of the low temperature ($\sim$10\,K) of the dust grains, and therefore another mechanism must exist for desorbing material from the grain surfaces. Hence several non-thermal desorption mechanisms have been proposed, such as the energy released in exothermic association reactions that leads to the immediate desorption of the reaction product \citep[e.g.,][]{Garrod07,Minissale16a}, or desorption via the absorption of FUV photons \citep[e.g.,][]{Hollenbach09}. The most widely adopted non-thermal desorption mechanism in gas-grain models describing the chemical evolution internal to dense clouds is cosmic ray-induced desorption (hereafter simply CRD). This mechanism is based on the notion that CRs sporadically strike grains, which are then transiently heated to a high temperature. The energy deposited and thus the transient maximum temperature reached by the grain depends on the energy of the incoming CR and on the grain properties, such as the stopping power of the grain material, and the grain size \citep[e.g.,][]{Leger85,Najita01}. The transient heating induces efficient desorption of the ice mantle surrounding the grain core, cooling the grain back down to its original equilibrium temperature. A numerical description of this process has been given by \citeauthor{Hasegawa93a}\,(\citeyear{Hasegawa93a}; hereafter HH93) in a form appropriate for application in rate-equation models, and this description has been subsequently adopted in many gas-grain chemical models. An alternative method, applied to Monte Carlo simulations of ice chemistry, has been presented by \citet{Herbst06}.

The cooling time of a transiently heated grain depends on the binding energy of the molecules participating in the cooling. The HH93 description of CRD is based on the assumption that the ice mantle consists of a homogeneous layer of a generic volatile molecule that has a binding energy comparable to that of CO, and therefore the grain cooling time is set to a constant value. Observations have indicated that water ice layers form on grains already at a visual extinction of a few magnitudes \citep{Whittet88}, and subsequent chemical evolution in denser gas leads to the deposition of layers of weakly polar molecules like CO in the top layers \citep{Tielens91}. Indeed, interstellar ices are expected to be heterogeneous and therefore the grain cooling time should be a time-dependent quantity.

The main aim of the present work is to provide a revised description of CRD that treats the process of grain cooling time-dependently, thus improving the accuracy of chemical simulations that consider CRD. This is achieved by considering quantities that are inherently part of any gas-grain chemical model -- the abundances of various molecules in the ice -- and hence the application of our revised description of CRD is straightforward also in models other than ours. We examine the transient heating process as well, although the main emphasis is presently on the cooling.

The paper is structured as follows. In Sect.\,\ref{s:theory}, we introduce the theoretical formulation of the revised CRD process, including transient heating and subsequent grain cooling, and discuss the different models that we consider in this paper. Section~\ref{s:results} presents the main results of our analysis, i.e., time-dependent chemical abundances in the gas phase calculated using the various models. We discuss some key points and the limitations of our current work in Sect.\,\ref{s:discussion}, and apply our new CRD scheme to a model of the prestellar core L1544. We give our concluding remarks in Sect.\,\ref{s:conclusions}. Some complementary simulation results are presented in the Appendix.

\section{Desorption of interstellar ices induced by cosmic rays}\label{s:theory}

The CRD model presented by HH93 assumes that the grain retains the transiently elevated temperature throughout the cooling process, i.e., the induced desorption events occur at constant temperature. The rate coefficient of such a desorption event for a molecule $i$ is then given by
\begin{equation}\label{k_HH93}
k_{\rm CR}(i) = f(a,T_{\rm max}) \, k_{\rm therm}(i,T_{\rm max}) \,
\end{equation}
where $a$ is the grain radius, $T_{\rm max}$ is the temperature that a grain of radius $a$ is heated to by a CR strike, and $k_{\rm therm}(i,T_{\rm max})$ is the thermal desorption rate coefficient of molecule $i$ at temperature $T_{\rm max}$. The pre-factor $f(a,T_{\rm max})$ is essentially an efficiency term (called ``duty cycle'' by HH93) determined by the ratio of the cooling time of the grains to the transient heating interval, which we here refer to as $\tau_{\rm cool}$ and $\tau_{\rm heat}$, respectively. In their model, HH93 assumed that CRs in the 20~to~70~MeV\,nucleon$^{-1}$ energy range deposit 0.4\,MeV of energy into 0.1\,$\mu$m grains; applying the heat capacity formula of \citet{Leger85} then leads to $T_{\rm max} \sim 70\,\rm K$. Assuming that CRs consist of iron nuclei and using the Fe CR flux from \citet{Leger85}, HH93 estimated a heating interval of $\tau_{\rm heat} = 3.16 \times 10^{13} \, \rm s$. They assumed that the grains are coated with a homogeneous ice layer made up of a generic CO-like molecule with a binding energy of 1200\,K. The desorption timescale of such a molecule at 70\,K is of the order of $10^{-5} \, \rm s$, which HH93 took to represent the grain cooling time. Therefore, for 0.1\,$\mu$m grains, $f(a,T_{\rm max}) = \tau_{\rm cool}/\tau_{\rm heat} = 3.16 \times 10^{-19}$. This value is customarily adopted in gas-grain chemical models that include CRD.

In the present work, we expand on the HH93 formulation of the CRD process by introducing a dynamic grain cooling mechanism that depends on time-dependent molecular abundances in the ice. We also modify the heating part by considering variations of the incident CR spectrum. Our new model is applicable to a range of grain sizes, but here we only discuss 0.1\,$\mu$m grains so that our results can be easily compared with those obtained using the HH93 description of CRD (for the same reason we also set $T_{\rm max} = 70 \, \rm K)$. The following describes the theoretical framework of the model in detail.

\subsection{Dynamic cooling}\label{ss:cooling}

The time-dependent grain cooling time is given by
\begin{equation}\label{eq:coolingTime}
\tau_{\rm cool} = \frac{E_{\rm th}}{\dot{E}} =  \frac{E_{\rm th}}{\dot{E}_{\rm subl} + \dot{E}_{\rm rad}} \, ,
\end{equation}
where $E_{\rm th}$ is the energy deposited into the grain by a CR strike (see Sect.\,\ref{ss:heating}), raising the grain temperature to $T_{\rm max}$. The rate of thermal energy loss, $\dot{E}$, is a sum of two terms: $\dot{E}_{\rm subl}$ and $\dot{E}_{\rm rad}$. The first term represents the desorption of ice via sublimation at $T = T_{\rm max}$, and is defined as
\begin{equation}\label{eq:evaporationRate}
\dot{E}_{\rm subl} = N_{\rm des} \sum_i k_{\rm B} E_b(i) \, \theta(i) \, \nu(i) \exp(-E_b(i) / T_{\rm max}) \, ,
\end{equation}
where the sum is over the atoms and molecules that contribute to the cooling; $E_b(i)$ is the binding energy in~K; $\nu(i)$ is the attempt frequency in~s$^{-1}$; $\theta(i)$ is the abundance fraction of species $i$ in the ice. The efficiency of the cooling depends strongly on $N_{\rm des}$, the total number of atoms and molecules available as coolants. We consider here two different approaches:
\begin{itemize}
\item[1)] A two-phase (gas+ice) chemical model, where the ice is treated as a single active layer. Hence the entire ice mantle is available for desorption, and $N_{\rm des} = \sum_i n(i)$, where the sum is over all atoms and molecules in the ice.
\item[2)] A three-phase chemical model, where the ice is separated into an active surface layer and an inert bulk (following \citealt{Hasegawa93b}), and hence $N_{\rm des} = \sum_j n(j)$, where the sum is now over atoms and molecules in the surface layer only.
\end{itemize}
Table~\ref{tab:coolants} lists the species that participate in the cooling, along with their binding energy (on a water ice surface). The entry marked ``others'' represents every other species present on the grain at any given time besides the ones that are explicitly included in Table~\ref{tab:coolants}. The binding energy taken for the ``other'' species is a parameter of the model and has been arbitrarily chosen, although its value influences our results only very little; we have found through testing that our results are only affected noticeably if the value is decreased to $\sim$1500\,K or below. This is a very unlikely scenario given that most of the molecules have a binding energy of several thousand~K. We have chosen a set of chemical species that are a priori expected to be present on a grain with appreciable abundances at various points during the accumulation of the ice. This ensures that the cooling time is calculated accurately at early times and at late times as the chemical simulation progresses.

\begin{table}[t]
	\centering
	\caption{Binding energies of the chemical species that participate in the cooling.}
	\begin{tabular}{cc}
		\hline
		\hline
		Species & Binding energy [K]\\
		\hline
		$\rm H$ & $450$\\
		$\rm C$ & $800$\\
		$\rm N$ & $800$\\
		$\rm CH$ & $925$\\
		$\rm N_2$ & $1000$\\
		$\rm CO$ & $1150$\\
		$\rm CH_4$ & $1300$\\
		$\rm O$ & $1390$\\
		$\rm oNH_3$ & $5534$\\
		$\rm pNH_3$ & $5534$\\
		$\rm oH_2O$ & $5700$\\
		$\rm pH_2O$ & $5700$\\
		$\rm others$ & $3000$\\
		\hline
	\end{tabular}
	\label{tab:coolants}
\end{table}

The second term in the denominator in Eq.\,(\ref{eq:coolingTime}) represents radiative cooling of the grain via blackbody emission at $T = T_{\rm max}$. This is given by \citep{Draine11}
\begin{equation}\label{eq:radiativeCooling}
\dot{E}_{\rm rad} = 4  \pi a^3 q_{\rm abs} \sigma T_{\rm max}^6 \, ,
\end{equation}
where $\sigma$ is the Stefan-Boltzmann constant and $q_{\rm abs} = 0.13 \, \rm K^{-2} \, cm^{-1}$ is a material-dependent numerical factor; the value of $q_{\rm abs}$ adopted here is valid for silicate grains. In practice, cooling via desorption dominates over radiative cooling for the model parameters considered in the present work, and our results are thus insensitive to the adopted value of $q_{\rm abs}$ (for carbonaceous grains, $q_{\rm abs} = 0.08 \, \rm K^{-2} \, cm^{-1}$; \citealt{Draine11}). We discuss radiative cooling briefly in Sect.\,\ref{ss:radiativeCooling}.

We assume, like HH93, that the sublimation rates of different species can be described in terms of the maximum grain temperature. In reality, the grain temperature changes as the ice sublimates (see for example Fig.\,9 in \citealt{Kalvans20}). A more precise treatment of the problem in a gas-grain chemical model such as ours would involve the integration of the desorption rate coefficients from $T_{\rm max}$ down to the equilibrium grain temperature. However, the present work aims to provide a description of CRD that is more accurate than the HH93 approach, but is still straightforward to apply to a gas-grain chemical model. Hence we leave a fully time-dependent study for future work.

\subsection{Transient grain heating}\label{ss:heating}

\begin{table*}
	\centering
	\caption{Values of the transient grain heating interval $\tau_{\rm heat}$ (in seconds) considered in this paper. The scaling factors used to obtain the Fe flux from the proton flux are indicated on the header row. The assumed CR energy range is 0.026 to 70~MeV~nucleon$^{-1}$.}
	\begin{tabular}{cccc}
		\hline
		\hline
		Visual extinction\,[mag]& P18H $\times \, 3.1 \times 10^{-4}$ & P18L $\times \, 3.1 \times 10^{-4}$ & L85 $\times \, 1.6 \times 10^{-4}$\\
		\hline
		0 & $1.34 \times 10^{11}$ & $4.84 \times 10^{12}$ & $3.35 \times 10^{13} \,^{(a)}$\\
		5 & $7.98 \times 10^{11}$ & $8.05 \times 10^{12}$ & $3.07 \times 10^{13}$\\
		10 & $1.15 \times 10^{12}$ & $9.72 \times 10^{12}$ & $3.21 \times 10^{13}$\\
		15 & $1.47 \times 10^{12}$ & $1.11 \times 10^{13}$ & $3.38 \times 10^{13}$\\
		30 & $2.40 \times 10^{12}$ & $1.47 \times 10^{13}$ & $3.97 \times 10^{13}$\\
		50 & $3.62 \times 10^{12}$ & $1.87 \times 10^{13}$ & $4.76 \times 10^{13}$\\
		100 & $6.80 \times 10^{12}$ & $2.74 \times 10^{13}$ & $6.39 \times 10^{13}$\\
		\hline
	\end{tabular}
	\tablenotetext{a}{The heating interval generally increases with increasing attenuation because low-energy CRs are removed from the relevant energy range faster than high-energy ones are added. Because the spectrum in L85 contains no particles with energy below 20\,MeV, a small attenuating column has the effect of adding higher-energy particles to the relevant energy range without removing the low-energy ones. For this reason, the heating interval is smaller for $A_{\rm V} = 5\,\rm mag$ than for $A_{\rm V} = 0\,\rm mag$ in the case with the L85 external spectrum.}
	\label{tab:Rheat}
\end{table*}

\begin{table}
	\centering
	\caption{Values of the cosmic-ray ionization rate $\zeta$ considered in this paper.}
	\begin{tabular}{cccc}
		\hline
		\hline
		Fe CR flux from & $\zeta \, \left[\rm s^{-1}\right]$ \\
		\hline
		\citet{Leger85} (L85) & $1.3 \times 10^{-17}$ \\
		\citet{Padovani18}, spectrum~$\mathscr{L}$ (P18L)  & $1.3 \times 10^{-17}$ \\
		\citet{Padovani18}, spectrum~$\mathscr{H}$ (P18H) & $1.0 \times 10^{-16}$ \\
		\hline
	\end{tabular}
	\label{tab:zeta}
\end{table}

As noted above, HH93 estimated a transient grain heating interval of $\tau_{\rm heat} = 3.16 \times 10^{13} \, \rm s$ based on the Fe CR flux given by \citet{Leger85}, corresponding to the 20 to 70~MeV~nucleon$^{-1}$ CR energy range. The value of $\tau_{\rm heat}$ is a function of several parameters, such as the shape of CR spectrum and the assumed CR energy range, and is also affected by other assumptions such as the composition of CRs themselves and whether the grains are coated in ice or not. Furthermore, the degree to which the CR flux is attenuated inside dense clouds depends both on the CR mass and the column density of the cloud.

In this paper, we do not consider CR species besides iron, so that our results are more directly comparable to HH93. However, we do present new values for the heating interval calculated using more modern estimates of the iron CR flux. We assume exterior to the cloud that the iron flux is proportional to the proton flux. To normalize the flux, we take the ratio of the iron flux from \citet{Ave08} at a kinetic energy of 0.9\,$\rm GeV \, nucleon^{-1}$ to the proton flux at a kinetic energy of 0.9\,GeV from \citet{Aguilar15}, resulting in an Fe/H ratio of $3.1 \times 10^{-4}$. To obtain the CR spectrum within the cloud, we use the propagation model described in \citeauthor{Padovani18}\,(\citeyear{Padovani18}; hereafter P18). We consider the CR flux given by \citeauthor{Leger85}\,(\citeyear{Leger85}; hereafter L85), as well as the fluxes recently presented by \citeauthor{Padovani18}\,(\citeyear{Padovani18}; their models $\mathscr{L}$ and $\mathscr{H}$; hereafter P18L and P18H, respectively). To derive heating intervals, we first used the attenuation model in \citet{Padovani18} to calculate the local spectrum of CRs at a range of visual extinctions from 0 to 100\,mag. In deriving heating intervals from these local spectra, we assumed that strikes from CRs in the energy range from 0.026 to 70 MeV raise the grain temperature to 70\,K. We note that the adopted CR energy range, 0.026 to 70~MeV~nucleon$^{-1}$, which accounts for low-energy CRs that are also able to deposit energy efficiently into the grains, differs from the energy range 20 to 70~MeV~nucleon$^{-1}$ assumed by HH93. We assume that the loss function for iron is equal to the loss function for hydrogen in \citet{Padovani18} times a factor of $Z^2 / A \approx 12$, where $Z$ is the atomic number and $A$ is the atomic mass number. The resulting values of $\tau_{\rm heat}$ are collected in Table~\ref{tab:Rheat}, where we also note the scaling factor used to obtain the Fe flux from the proton flux. In the case of L85, we used a slightly lower scaling factor ($1.6 \times 10^{-4}$) in order to maintain consistency with the work of HH93. Our value for $\tau_{\rm heat}$ using the L85 CR spectrum is very close to the value estimated by HH93 (assuming $A_{\rm V} = 10 \, \rm mag$). The \citet{Padovani18} Fe fluxes are on the other hand much higher, which has an effect on gas-phase chemistry as it will be evident below. We also note that our $\tau_{\rm heat}$ values are a factor of several lower than those calculated by \citet{Kalvans18b}. The reason for this discrepancy is at present unknown, and will not be addressed in the present work.

Variations in the CR flux affect the CR ionization rate, which also depends on the column density, i.e., depth into the cloud. However, the value of the CR ionization rate also has a large effect on the chemistry, and therefore varying the ionization rate from model to model would make it difficult to disentangle the effect of the dynamic cooling alone on the simulation results. We therefore adopt a simple set of CR ionization rate values, collected in Table~\ref{tab:zeta}, in most of our models (see below), depending on the CR flux. We note that using the P18H CR flux leads to a notably higher ionization rate than in the other two cases.

The derivation of the transient maximum grain temperature in HH93, based on the work of \citet{Leger85}, ignores the effect of the ice mantle that accumulates on the grain over time. As the ice thickness increases, the energy-absorbing volume increases and the value of $T_{\rm max}$ decreases. This issue is particularly important for small grains for which the relative volume increase due to mantle growth is substantial. Furthermore, a spectrum of CR energies and impact parameters leads to a distribution of $T_{\rm max}$ values instead of a single, unique value for a given grain size. For simplicity and to ensure that our results can be meaningfully compared with the HH93 description of CRD, we adopt the same amount of energy deposited by a CR strike into a 0.1\,$\mu$m grain (0.4~MeV) as HH93 did. It should be pointed out that the exact value of the deposited energy depends on the CR energy, the composition of the grain, and the impact parameter. We do not consider variations of the grain radius in this paper.

We will present a time-dependent transient grain heating model in an upcoming work (Sipil\"a et al., in prep.) that will include not only the effect of ice mantle growth in decreasing $T_{\rm max}$, but also the effect of considering a spectrum of $T_{\rm max}$ values for a given grain size. The basis for that model is a new detailed study of the grain-size dependence of ice mantle thickness that will be presented elsewhere (Silsbee et al., in prep.).

\subsection{Final form of the CRD rate coefficient}\label{ss:finalRateCoeff}

To summarize the previous subsections, the significant differences between the present work and that of HH93 are that we calculate the grain cooling time (set to constant $10^{-5}\,\rm s$ in HH93) dynamically based on time-dependent ice abundances, and consider variations of the CR flux which impacts the grain heating interval (set to constant $3.16 \times 10^{13}\,\rm s$ in HH93). Hence in our formulation of CRD, the constant $f(70\,\rm K)$ term of HH93 (cf. their Eq.\,15) is replaced by the more general $f(a,T_{\rm max}) = \tau_{\rm cool}/\tau_{\rm heat}$, where $\tau_{\rm cool}$ is calculated from Eqs.\,(\ref{eq:coolingTime})~to~(\ref{eq:radiativeCooling}) and $\tau_{\rm heat}$ can be either read or interpolated from Table~\ref{tab:Rheat}.

One further issue must be taken into account to ensure that our new method is applicable in a wide variety of physical conditions. It is possible that the number of volatile species, which are the ones that efficiently remove heat from the grains, are not numerous enough in the ice to be able to shed all of the energy deposited by the impinging CR. In such a case, only a fraction of the deposited energy is removed via sublimation, and the remaining heat is lost via radiation. Desorption rates of volatiles in particular will be overestimated if one adopts in such situations the $f(a,T_{\rm max})$ term as is, because the lack of efficient coolants will be compensated by an increase in the grain cooling time, leading to an increased CRD rate for all molecules. The frequency of desorption events is naturally limited by the frequency of CR strikes -- therefore, we impose an upper limit of $\tau_{\rm heat}^{-1}$ to the desorption rate of any molecule, yielding the final form of the CRD rate coefficient:
\begin{equation}\label{kFinal}
k_{\rm CR}(i) = \min \left( f(a,T_{\rm max}) \, k_{\rm therm}(i,T_{\rm max}) , \tau_{\rm heat}^{-1} \right) .
\end{equation}
This formulation ensures that desorption rates are not overestimated if the total abundance of volatiles in the ice is low. The rate coefficient as given in Eq.\,(\ref{kFinal}) is applied to all species in the model, {\sl except $\rm H_2$ (and its deuterated isotopologs), for which we apply Eq.\,(\ref{k_HH93}).} This is because $\rm H_2$ is not allowed to participate in the cooling, and limiting its desorption leads to other issues; see discussion in Sects.\,\ref{ss:H2Cooling}~and~\ref{ss:L1544}. Because of its very low binding energy, $\rm H_2$ can be desorbed also by much more frequent strikes from CR protons which do not elevate the grain temperature sufficiently to desorb significant amounts of other molecules.  For this reason, it is inappropriate to limit its desorption rate to the frequency of impacts of iron nuclei. Our new formulation of CRD can thus be implemented in a gas-grain chemical model by employing Eq.\,(\ref{kFinal}) for all species except $\rm H_2$, as opposed to using the formulation of HH93 which is based on a constant grain cooling time.

We reiterate that in this paper we set $a = 0.1\,\rm \mu m$ and $T_{\rm max} = 70 \, \rm K$ to facilitate the comparison with HH93; for other values of these parameters, the heating intervals need to be recalculated. We are currently working on a more general CRD model that relaxes the assumptions on $a$ and $T_{\rm max}$; the model will be presented in an upcoming dedicated paper.

\subsection{Chemical and physical model}\label{ss:chemicalModel}

\begin{table*}
	\centering
	\caption{Values of the fiducial physical model parameters and descriptions of the four model cases considered here.}
	\begin{tabular}{cc}
		\hline
		\hline
		Physical model parameters & Value \\
		\hline
		Volume density of the medium $n({\rm H_2})~[{\rm cm^{-3}}]$ & $10^3$, $10^5$, $10^6$\\
		Gas temperature $T_{\rm gas}~[{\rm K}]$ & $15$, $10$, $10$\\
		Dust temperature $T_{\rm dust}~[{\rm K}]$ & $15$, $10$, $10$\\
		Visual extinction $A_{\rm V}~[{\rm mag}]$ & $2$, $10$, $30$\\
		Grain radius $a~[{\rm \mu m}]$ & $0.1$\\
		\hline
		\hline
		Model & Description\\
		\hline
		F2 & Fiducial model, two-phase ice chemistry, grain cooling following HH93\\
		D2 & As F2, but with dynamic grain cooling\\
		F3 & Fiducial model, three-phase ice chemistry, grain cooling following HH93\\
		D3 & As F3, but with dynamic grain cooling\\
		\hline
	\end{tabular}
	\label{tab:models}
\end{table*}

We apply the new description of CRD to our gas-grain chemical model, which simulates chemical evolution by the rate-equation method. The basic functionality of the model, including the equations describing chemical reactions in the gas phase and on the grain surfaces as well as the gas-grain chemical interaction itself, is laid out in \citet{Sipila15a}. A very important detail in the present context is that the model tracks the time-dependent changes in molecular abundances not only in the gas phase but also on the grain surfaces, which is necessary for the correct estimation of the sublimation rate (Eq.\,\ref{eq:evaporationRate}). We employ our full chemical networks including deuterium and spin-state chemistry \citep{Sipila15b,Sipila19b}. The chemical network contains a combined total of $\sim$75700 reactions, $\sim$2100 of which are grain-surface reactions. We adopt the initial elemental abundances given in Table~1 in \citet{Sipila19b}. We note that deuterated species are included in the chemical network so that we can assess the effect of our new CRD model on deuteration, which is important from an observational point of view. Our main conclusions regarding for example the grain cooling time would not be affected if the chemical network was simplified to include only non-deuterated molecules.

To reduce the number of model parameters and to simplify the interpretation of our results, we adopt a set of zero-dimensional physical models to run our chemical simulations. The fiducial model parameters are given in Table~\ref{tab:models}. We consider physical conditions appropriate to the centers of starless and prestellar cores, as well as lower-density gas associated with the envelopes surrounding the cores. Sub-7\,K temperatures have been inferred deep inside starless and prestellar cores \citep[e.g.,][]{Crapsi07,Pagani07,Harju08}, and thus our chosen values for the gas and dust temperatures in the high-density models are not completely consistent with high-density gas in low-mass star-forming regions. Our choice of 10\,K for both high-density models is simply for the sake of easing the interpretation of our results. An increase of $A_{\rm V}$ from 10 to 30\,mag has a negligible effect on photochemistry, but does increase $\tau_{\rm heat}$ by a small factor (Table~\ref{tab:Rheat}). We set the gas temperature to 15\,K and $A_{\rm V}$ to 2\,mag in the low-density model, mimicking physical conditions typical to molecular cloud envelopes. We include in all simulations the effect of $\rm H_2$ self-shielding \citep{Draine96}, which is important at low volume densities as it prevents the excess formation of atomic hydrogen from $\rm H_2$ dissociation. For this, we derived the $\rm H_2$ column density using $N({\rm H_2})/A_{\rm V} = 1.2 \times 10^{21} \, \rm cm^{-2}$. Self-shielding of other molecules, such as CO or $\rm N_2$, is not considered in the present simulations.

As noted in Sect.\,\ref{ss:cooling}, we study two-phase and three-phase chemical models, and include in these models either the HH93 description of CRD or the revised, dynamic one presented in this work. This leads to four distinct models, also indicated in Table~\ref{tab:models}. Models~F2~and~F3 refer to two-phase and three-phase chemical models, respectively, adopting the HH93 CRD description. In models~D2~and~D3 we adopt instead our new, dynamic approach to CRD.

In addition to the zero-dimensional physical models, we have explored the effect of dynamic CRD on chemical evolution in the prestellar core L1544. These simulations are described in Sect.\,\ref{ss:L1544}.

\section{Results}\label{s:results}

In what follows, we present the evolution of the grain cooling time as well as gas-phase and ice abundances as a function of simulation time corresponding to the four models collected in Table~\ref{tab:models}, for the L85 CR spectrum and the P18L CR spectrum. Corresponding results for the P18H CR spectrum are presented in the Appendix. We emphasize that in our simulations we use absolute numbers calculated from the local spectra (cf. Tables~\ref{tab:Rheat}~and~\ref{tab:zeta}) and not the spectra themselves. However we refer to the different cases as ``spectra'' to differentiate between input data.

\begin{figure*}
	\centering
	\includegraphics[width=2.0\columnwidth]{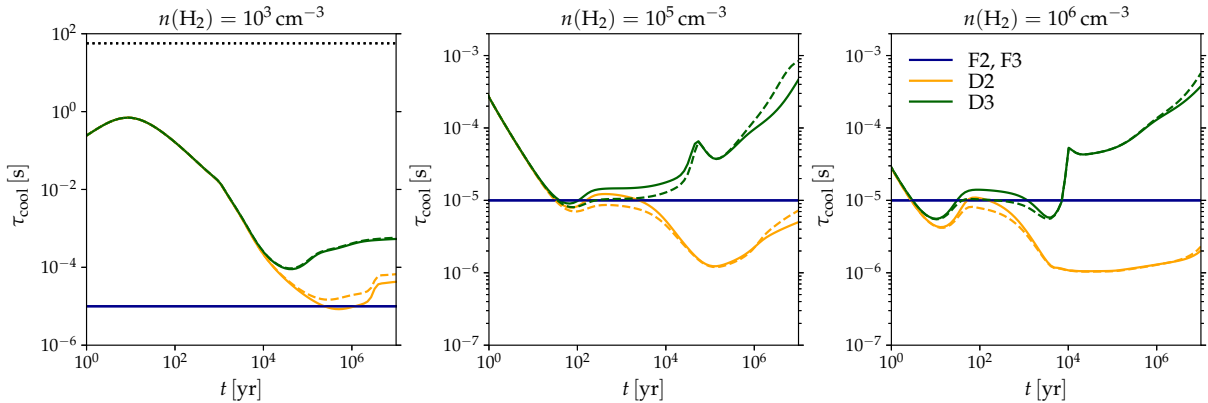}
    \caption{Grain cooling time $\tau_{\rm cool}$ as a function of simulation time for models F2, D2, F3, and D3, assuming $n({\rm H_2}) = 10^3 \, \rm cm^{-3}$ ({\sl left}), $n({\rm H_2}) = 10^5 \, \rm cm^{-3}$ ({\sl middle}), or $n({\rm H_2}) = 10^6 \, \rm cm^{-3}$ ({\sl right}). {\sl Solid} lines correspond to models adopting the L85 CR spectrum, while {\sl dashed} lines correspond to models adopting the P18L CR spectrum. Note the difference in y-axis range between the panels. The radiative grain cooling time is displayed as the dotted black line in the left-hand panel.}
    \label{fig:coolingTime}
\end{figure*}

\subsection{Grain cooling time}

Figure~\ref{fig:coolingTime} shows the time-evolution of $\tau_{\rm cool}$ for models F2, D2, F3, and D3 at the three volume densities. The grain cooling time is equal in models F2 and F3 which both adopt the HH93 description of CRD. Dynamic CRD affects the simulation results in different ways depending on the model parameters and also on the time. Let us first concentrate on the two high-density models. At early times when there is little ice on the grains, the grain cooling time predicted by our chemical model is very similar to that adopted by HH93, up to $t \sim 10^4 \, \rm yr$ at $n({\rm H_2}) = 10^5 \, \rm cm^{-3}$ and up to $t \sim 10^3 \, \rm yr$ at $n({\rm H_2}) = 10^6 \, \rm cm^{-3}$. Afterwards, the cooling time begins to diverge from the constant HH93 value – the two-phase and three-phase models evolve in different directions. In model~D2, the grain cooling time tends to decrease with time, because CRD can in this case occur from the entire ice mantle where weakly-bound, volatile molecules are abundant. Allowing desorption from the top layer only (model~D3) leads to an eventual increase in the grain cooling time due to the trapping of volatiles in the inert mantle; at late times the active surface layer is mostly composed of strongly-bound species, and there are not enough volatiles in the surface layer to remove all of the heat deposited in a CR strike. The time-evolution of the cooling times is qualitatively similar for $n({\rm H_2}) = 10^5 \, \rm cm^{-3}$ and $10^6 \, \rm cm^{-3}$. Switching between the two CR spectra has only a minor influence on the cooling time, despite the fact that a change in $\tau_{\rm heat}$ affects desorption rates and hence the ice chemistry. When adopting the P18L CR spectrum, the more frequent CR strikes as compared to L85 lead to more efficient desorption (i.e., a larger $f(a,T_{\rm max})$ factor) causing the abundances of volatiles to drop, and hence the grain cooling time increases slightly at late simulation times.

The grain cooling time in physical conditions corresponding to molecular cloud envelopes is, at early simulation times, clearly higher than $10^{-5} \, \rm s$. This is because in the beginning of the simulation there is very little ice, and hence only a small population of volatiles, on the grains and only a fraction of the deposited heat can be removed via sublimation. A similar divergence of the two-phase and three-phase model results compared to the high-density cases manifests itself, but naturally this occurs only at later simulation times when more than one monolayer of ice has formed on the grains. Also here the effect of changing the CR spectrum is minor.

Figure~\ref{fig:coolingTime} also shows that the sublimation timescale never reaches such high values so as to (near) equalize the grain cooling time (including sublimation and radiative cooling) and the radiative cooling time.

\subsection{Gas-phase molecular abundances}

\begin{figure*}
	\centering
	\includegraphics[width=1.9\columnwidth]{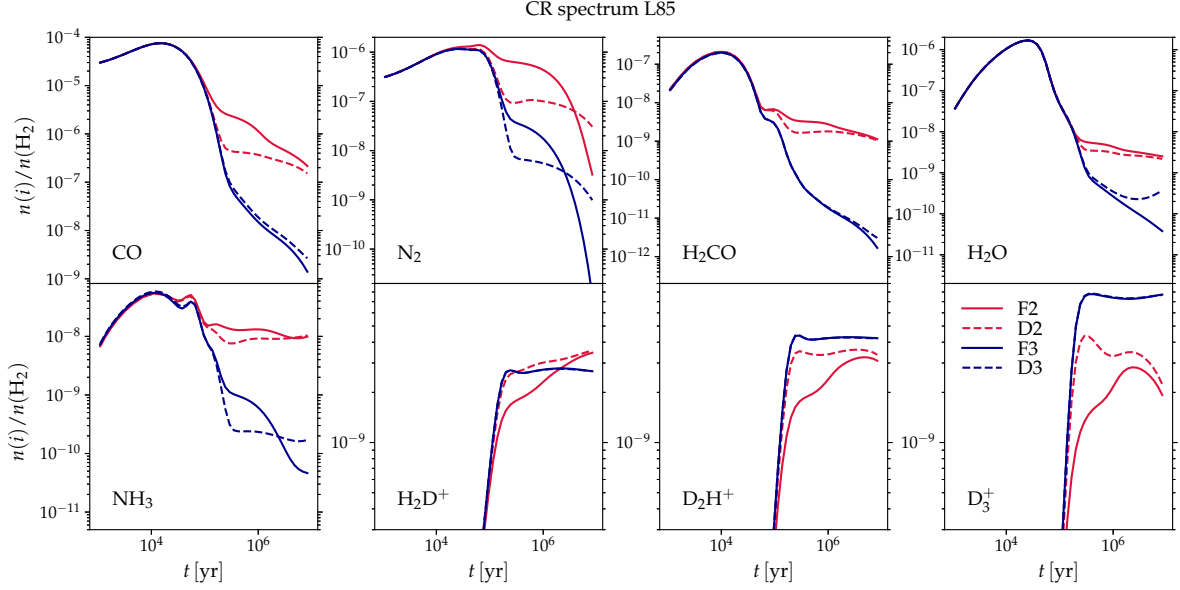}
    \caption{Abundances of selected gas-phase species as a function of time for models F2, D2, F3, and D3, labeled in the bottom right panel, assuming $n({\rm H_2}) = 10^5 \, \rm cm^{-3}$. Here, $\tau_{\rm heat}$ is calculated using the L85 CR spectrum.}
    \label{fig:abundances_1e5_Hasegawa93}
\end{figure*}

The differences in grain cooling time are naturally reflected in the molecular abundances. Let us again begin our analysis with the high-density physical conditions. Figure~\ref{fig:abundances_1e5_Hasegawa93} shows the time evolution of selected gas-phase molecular abundances for models F2, D2, F3, and D3 at $n({\rm H_2}) = 10^5 \, \rm cm^{-3}$ using the L85 CR spectrum. First, comparing the two-phase models F2 and D2 reveals that the introduction of dynamic CRD decreases the abundances of non-deuterium-bearing species after the onset of freeze-out by a varying degree, in most cases by a factor of 2-3. This is directly related to the temporal decrease of the grain cooling time (Fig.\,\ref{fig:coolingTime}); faster cooling of the grains means that all molecules that are not very lightly bound have little time to be desorbed. Consequently the overall degree of freeze-out is higher in model~D2 compared to F2, which also leads to the associated increase in the abundances of the main deuterium-carrying ions (linked to the CO depletion factor).

Switching from the two-phase model with constant grain cooling time (F2) to the corresponding three-phase model (F3) results in a drastic general drop of gas-phase abundances for non-deuterium-bearing species due to the trapping of molecules in the inert mantle, leading in particular to a catastrophic freeze-out of CO. In the three-phase models only the surface layer is active and available for desorption, and is composed of a mixture of volatiles and strongly-bound molecules. When using our dynamic description of CRD, the grain cooling time is in these conditions higher than $10^{-5}\,\rm s$ which results in a slight boost to the abundances of various molecules, but the effect is small because desorption rates are limited by the frequency of CR strikes. $\rm N_2$ represents an interesting counterexample, as its abundance decreases in model~D3 as compared to F3 after the onset of freeze-out. The latter model overestimates the desorption efficiency of $\rm N_2$ by about a factor of six\footnote{The ratio of CR strike frequency to $\rm N_2$ desorption rate coefficient is about six in model~F3.}, and the desorbed $\rm N_2$ is processed in the gas phase into nitrogen hydrides, and partly into atomic N which forms more hydrides upon adsorption onto grains. In model~D3, $\rm N_2$ is not desorbed as easily which limits the reprocessing of nitrogen, as icy $\rm N_2$ is extremely unreactive. This result highlights a shortcoming of the HH93 model in the overestimation of the CRD rates of volatiles with a binding energy lower than that of CO.

\begin{figure*}
	\centering
	\includegraphics[width=1.9\columnwidth]{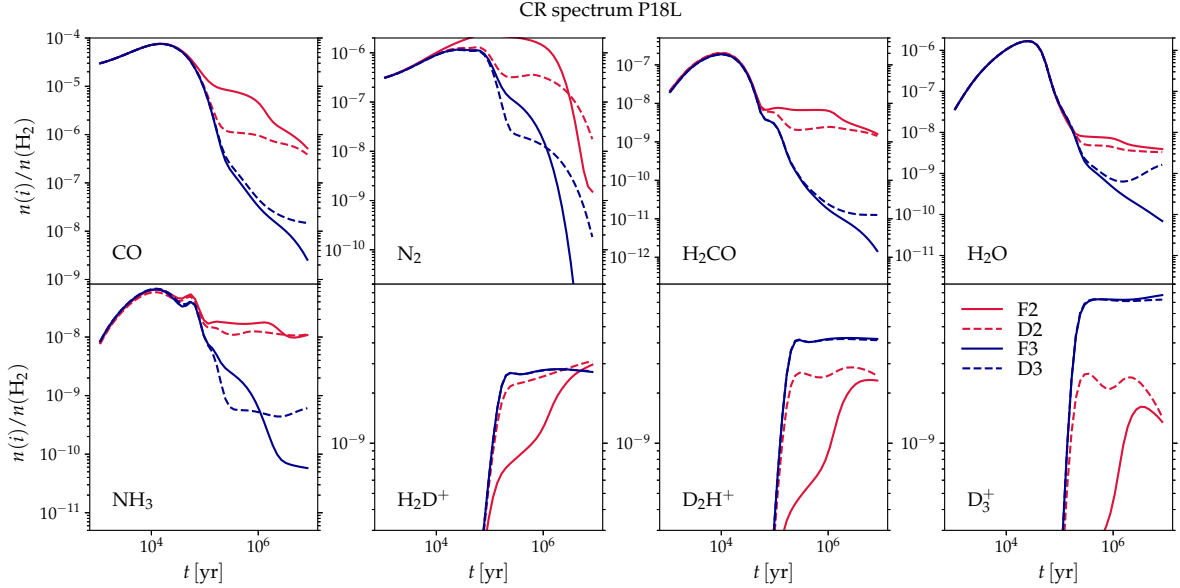}
    \caption{Abundances of selected gas-phase species as a function of time for models F2, D2, F3, and D3, labeled in the bottom right panel, assuming $n({\rm H_2}) = 10^5 \, \rm cm^{-3}$. Here, $\tau_{\rm heat}$ is calculated using the P18L CR spectrum, and the y-axis scale is the same as in the corresponding panels in Fig.\,\ref{fig:abundances_1e5_Hasegawa93} for the sake of easier comparison.}
    \label{fig:abundances_1e5_Padovani18Low}
\end{figure*}

When using the P18L CR spectrum (Fig.\,\ref{fig:abundances_1e5_Padovani18Low}), in which the heating events occur more often than in the case of the L85 spectrum, the results are very similar to those presented in Fig.\,\ref{fig:abundances_1e5_Hasegawa93}. This is expected based on the small differences in the grain cooling time (Fig.\,\ref{fig:coolingTime}). The two-phase model is more sensitive to changes in the heating timescale.

\begin{figure*}
	\centering
	\includegraphics[width=1.9\columnwidth]{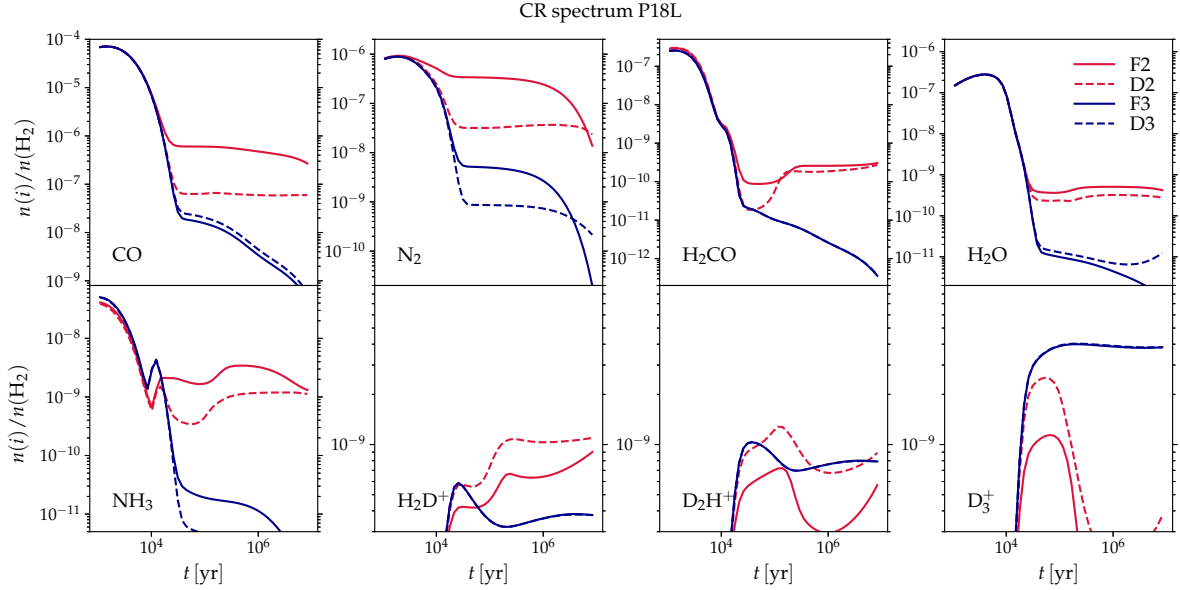}
    \caption{As Fig.\,\ref{fig:abundances_1e5_Padovani18Low}, but assuming $n({\rm H_2}) = 10^6 \, \rm cm^{-3}$.}
    \label{fig:abundances_1e6_Padovani18Low}
\end{figure*}

Figure~\ref{fig:abundances_1e6_Padovani18Low} presents the results of simulations assuming $n({\rm H_2}) = 10^6 \, \rm cm^{-3}$ and using the P18L CR spectrum. We default from here on to this spectrum, which is based on data from Voyager~1 \citep{Cummings16}. Comparing Figs.\,\ref{fig:abundances_1e5_Padovani18Low}~and~\ref{fig:abundances_1e6_Padovani18Low} reveals that the results at the two densities are qualitatively similar, but some important variation is introduced by the switch to higher densities. In general, the difference between the abundances predicted by models~D2~and~F2 increases with volume density. Curiously, the abundances of the main deuterated ions as predicted by models~F2~and~D2 are closer to each other at $n({\rm H_2}) = 10^6 \, \rm cm^{-3}$ than they are at $n({\rm H_2}) = 10^5 \, \rm cm^{-3}$ before $t \sim 10^6 \, \rm yr$, despite the larger CO depletion factor predicted by model~D2 at high density (the time-evolution of the $\rm H_2$ ortho/para ratio is virtually unaffected by the inclusion or exclusion of dynamic CRD). The difference between the results of models F3~and~D3 remains small also at high volume density.

\begin{figure*}
	\centering
	\includegraphics[width=1.95\columnwidth]{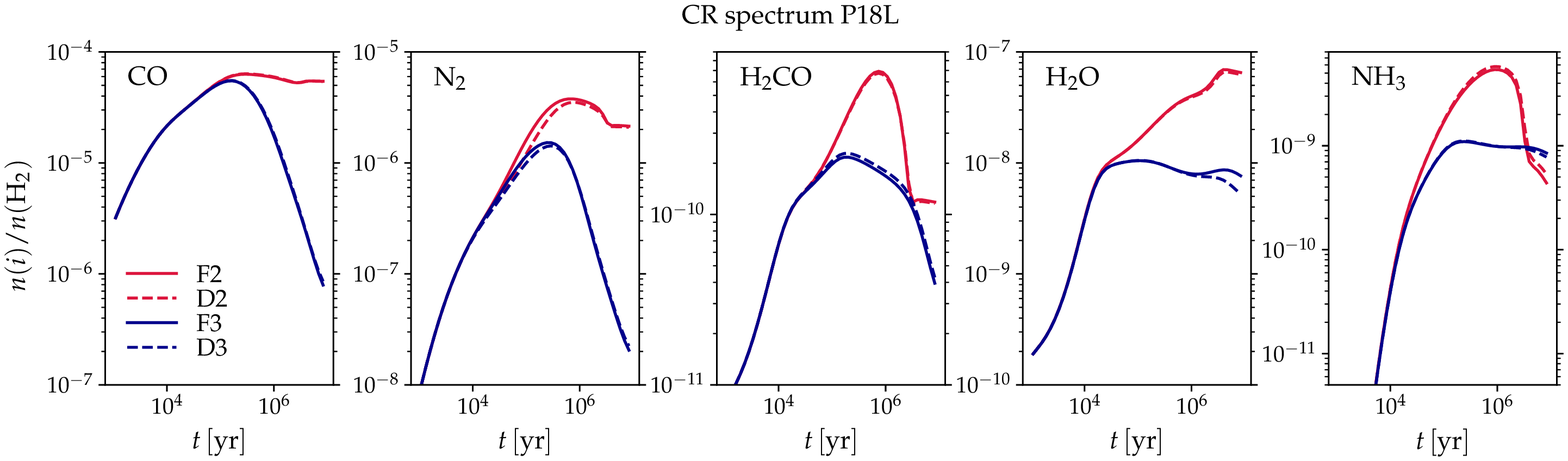}
    \caption{As Fig.\,\ref{fig:abundances_1e5_Padovani18Low}, but assuming $n({\rm H_2}) = 10^3 \, \rm cm^{-3}$. Deuterated ions are omitted owing to their very low abundances.}
    \label{fig:abundances_1e3_Padovani18Low}
\end{figure*}

In molecular cloud envelopes, one would a priori expect gas-phase chemical evolution to be dominated by photoionization and photodissociation, with only a minor role assigned to CRD owing to the low volume density and low visual extinction. This is confirmed by Fig.\,\ref{fig:abundances_1e3_Padovani18Low}, which shows that for low volume densities and hence for low total ice abundances, dynamic CRD has only a very minor effect despite the long grain cooling time. In low-density and poorly shielded conditions, such as in our molecular cloud envelope model, ice abundances remain low over long periods of time, and the ice is mostly made up of water with only minor contributions from volatile species such as CO. Water has a very high binding energy, and so it is very hard to sublimate in appreciable quantities through CRD. This translates to a long grain cooling time (Fig.\,\ref{fig:coolingTime}), because the cooling is dominated by radiation as opposed to sublimation. In more dense and better shielded physical conditions, such as cores within molecular clouds, the ice composition is much more varied with large contributions from relatively weakly bound volatiles (especially in the top ice layers), and the efficiency of CRD is dictated mainly by the desorption of these species instead of strongly-bound molecules like water. Our new CRD description captures these differences naturally, and describes the desorption process well in a wide variety of physical conditions (within the limitations of rate-equation models; see Sect.\,\ref{ss:diffusion}). This is in stark contrast with the HH93 model for CRD, which is appropriate only when the ice is made up of CO.

To summarize the results presented so far, we find that calculating the grain cooling time-dependently in physical conditions corresponding to the central areas of starless and prestellar cores decreases in general the various gas-phase abundances when one adopts a two-phase gas-grain model where the ice on the grain surface is essentially treated as one active layer. For three-phase models that separate the ice into an active surface layer on top of an inert mantle, our revised CRD model predicts generally very small differences compared to the case of constant grain cooling time, when only a single surface layer is considered active. {\sl We emphasize that this result will not hold anymore if multiple reactive layers are considered instead.} For physical conditions corresponding to molecular cloud envelopes, dynamic CRD has only a very limited effect on simulation results.

In the above, we concentrated exclusively on gas-phase molecules; we discuss the effects on ice chemistry in the next section.

\section{Discussion}\label{s:discussion}

The revised CRD scheme presented above is in our opinion a significant improvement over the time-independent one of HH93. However, there remain several issues that are potentially important but are not presently included in our model. In what follows, we describe these in some detail, and present some additional simulation results pertaining to ice abundances and to the prestellar core L1544.

\subsection{Contribution of $\rm H_2$ to grain cooling}\label{ss:H2Cooling}

Gas-grain chemical models often predict a high fraction of molecular hydrogen in the ice, which requires special treatment in a three-phase chemical model to avoid excessive amounts of $\rm H_2$ in the ice \citep{Hasegawa93b}. $\rm H_2$ is a non-polar molecule, and consequently the binding energies of molecules on $\rm H_2$ ice are much lower than on water ice, for example. Some numerical models have been presented in the literature that modify binding energies dynamically as a function of the $\rm H_2$ ice fraction \citep[e.g.,][]{Garrod11,Taquet14,Furuya15}. Such schemes result in a strong self-limiting of the surface $\rm H_2$ abundance because the binding energy of $\rm H_2$ on $\rm H_2$ is very low (45\,K; \citealt{Vidali91}), but these schemes are poorly constrained owing to the lack of experimental data on the subject.

$\rm H_2$ is very abundant, and could thus contribute appreciably to the grain cooling. We tested this by adding $\rm H_2$ to the list of cooling molecules (Table~\ref{tab:coolants}) and repeating the high-density cloud simulations presented above -- without considering dynamically-varying binding energies. In these tests, the grain cooling time dropped to $10^{-9}$\,s or even lower depending on the model parameters. This in turn led to an extreme (several orders of magnitude) gas-phase depletion factor at high densities, which is in contradiction with observations \citep[e.g.,][]{Caselli99}. It is clear that including the effect of $\rm H_2$ desorption on the grain cooling requires additional considerations for the model to produce meaningful results. One possibility is to employ dynamical binding energies, and we have also run some separate tests on this; however, doing so changes the results across all models, including the fiducial models using the HH93 CRD scheme. We leave a more detailed investigation of this topic to future work.

\subsection{Enhancement of diffusion and reaction rates; limitations of rate-equation models}\label{ss:diffusion}

\begin{figure*}
	\centering
	\includegraphics[width=1.9\columnwidth]{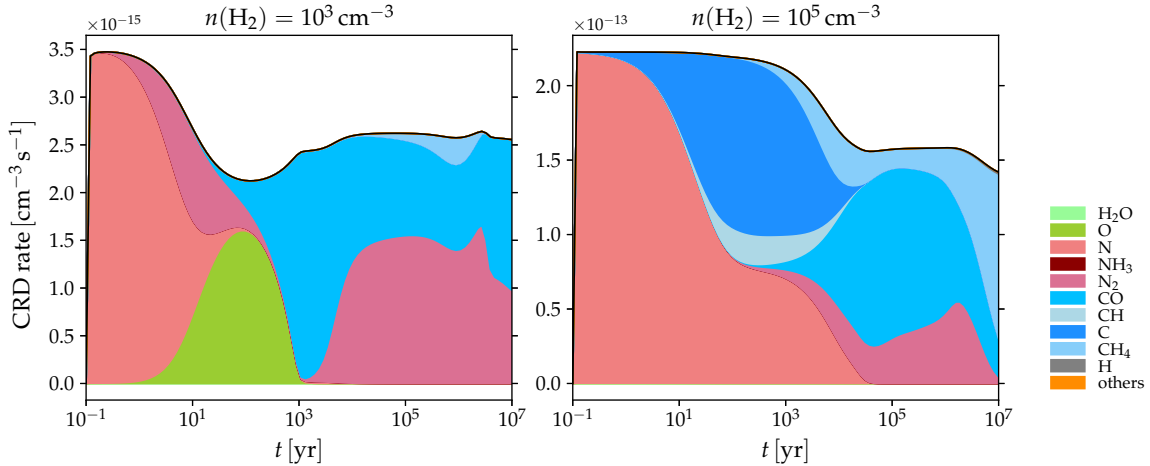}
    \caption{Total CRD rate of the species participating in the grain cooling ({\sl black solid} line) as a function of time in model~D2 at two different densities. The colors show the contributions of individual species to the total (see legend on the right). The species are grouped in color according to their elemental content.}
    \label{fig:desorptionRates}
\end{figure*}

The heating of a grain following a CR strike leads to the transient enhancement of the diffusion rates of molecules in the ice and allows molecules to more easily cross activation barriers, thereby increasing reaction rates as well. Indeed, recent simulations by \citet{Kalvans20} show that the enhanced diffusion plays a role in removing part of the deposited energy via the transport of molecules to the surface where they can be desorbed. This assumes that diffusion can occur outward from monolayers beneath the ice surface. The two-phase chemical model employed here contains the implicit assumption that diffusion can occur in all monolayers in all directions with the same efficiency, while in the three-phase model, diffusion beneath the surface layer is prohibited -- only surface-to-mantle exchange is allowed. We point out that reactions within the ice mantle have also been investigated in the context of three-phase models (e.g., \citealt{Garrod13b}; but see \citealt{Ghesquiere18} and \citealt{Shingledecker19} for evidence of non-diffusivity in the mantle). Recent experimental data suggest that, if diffusion does occur in the mantle, it does so along the surfaces of cracks or pores in the ice \citep{Tsuge20} that may or may not be present in the ice depending on its formation mechanism; water ice formed in situ, for example, is expected to be compact and amorphous \citep{Oba09}.

Our chemical model resolves the chemical evolution by solving a set of rate equations. Such methods are inherently poorly suited to modeling transient phenomena on grain surfaces. While adaptations can be made to describe transient effects \citep[e.g.,][]{Shingledecker18}, we have here refrained from making any modifications to the rate equation scheme so that our results can be compared to earlier models adopting the HH93 description of CRD. Furthermore, the work of \citet{Kalvans20} indicates that the effect of large-scale diffusion on the grain cooling is limited. Detailed dedicated simulations of the chemistry in the ice are required to ascertain whether transiently enhanced reaction rates could alter the large-scale ice content in a way that would also significantly affect the grain cooling timescale.

As discussed in Sect.\,\ref{ss:finalRateCoeff}, it is possible that the required number of molecules to dissipate the energy deposited in a CR strike exceeds the total number of molecules in the surface layer, and this applies in particular to volatiles as they are the principal coolants. Such situations can arise in low-density conditions, or in the three-phase chemical model (in all physical conditions) where only the surface layer is active. The limitations of rate-equation models in describing transient phenomena were already mentioned above -- this extends also to CRD events, because rate equations cannot account for the fact that molecules are lost during a single heating event. The desorption efficiencies at the start of a heating event are set by the ice composition at that time in the simulation, and are not assumed to vary as the grain cools. It is therefore necessary to impose an upper limit to desorption rates, which we here set to the frequency of CR strikes, to make sure that volatiles in particular are not desorbed at unrealistically high rates. We note that in two-phase chemical models the number of molecules available for desorption almost always greatly exceeds the required number to dissipate the deposited energy, and hence the upper limit for the desorption rate is in most cases not reached.

Finally, as evidence that it is indeed the desorption of volatiles that dominates the energy dissipation via sublimation, we show in Fig.\,\ref{fig:desorptionRates} the contributions of individual species to the total CRD rate in model~D2 at two different densities. It is clear that the contribution of strongly-bound molecules is negligible -- the desorption is dominated at late times by $\rm N_2$ and CO, and at high density by $\rm CH_4$ as well.

\subsection{Ice abundances}

\begin{figure*}
	\centering
	\includegraphics[width=1.9\columnwidth]{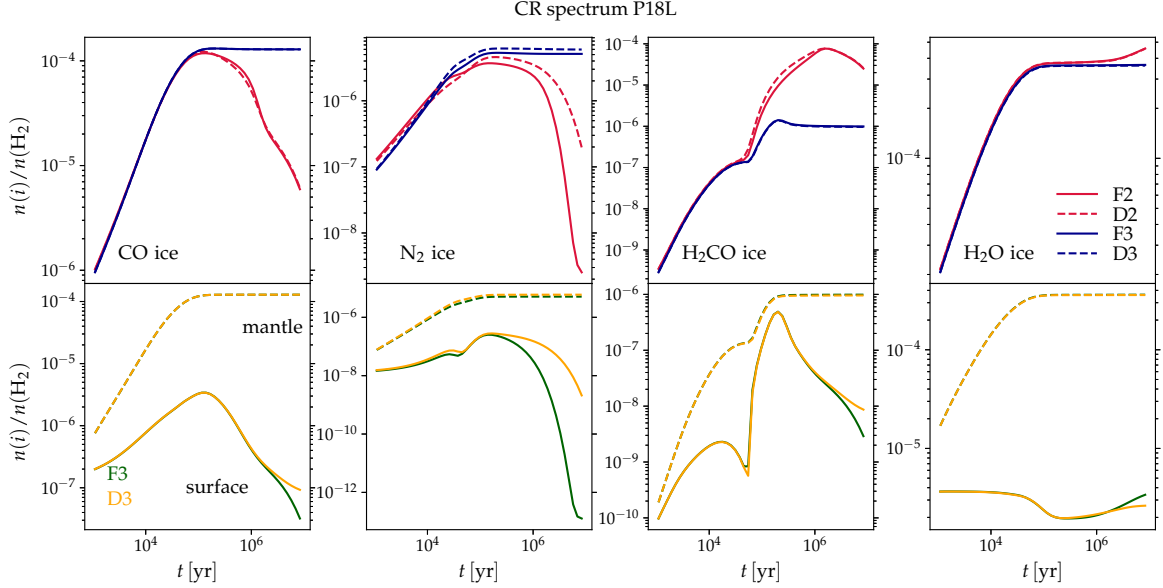}
    \caption{{\sl Top:} Abundances of selected ice molecules as a function of time for models F2, D2, F3, and D3, labeled in the top right panel. The abundances corresponding to the three-phase models F3 and D3 represent sums over the surface and mantle populations. {\sl Bottom:} Breakdown of the ice populations in the three-phase models F3 ({\sl green}) and D3 ({\sl orange}) into the active surface layer ({\sl solid} lines) and the inert mantle ({\sl dashed} lines), as labeled in the lower left panel. Here $n({\rm H_2}) = 10^5 \, \rm cm^{-3}$, and $\tau_{\rm heat}$ is calculated using the P18L CR spectrum.}
    \label{fig:abundances_1e5_ice_Padovani18Low}
\end{figure*}

\begin{figure*}
	\centering
	\includegraphics[width=1.9\columnwidth]{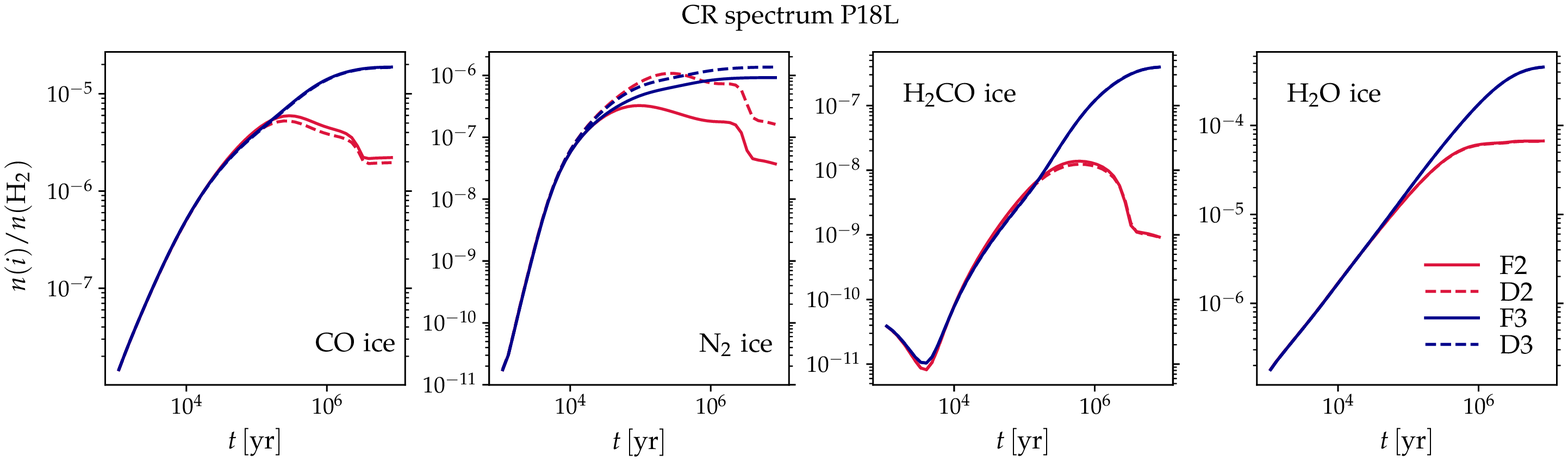}
    \caption{As Fig.\,\ref{fig:abundances_1e5_ice_Padovani18Low}, but assuming $n({\rm H_2}) = 10^3 \, \rm cm^{-3}$ and showing only the total ice abundances.}
    \label{fig:abundances_1e3_ice_Padovani18Low}
\end{figure*}

It is clear that the changes in gas-phase abundances resulting from the switch from static to time-dependent grain cooling should be reflected in the ice content as well. However, it turns out that the ice abundances are in general modified only very slightly from model to model, as displayed in Figs.\,\ref{fig:abundances_1e5_ice_Padovani18Low} and \ref{fig:abundances_1e3_ice_Padovani18Low}. The figures show that the difference between the F2 and D2 cases, and that between F3 and D3, is in general very small. Fundamental differences between the two-phase and three-phase models are also evident, such as the hydrogenation of CO ice which occurs only to a limited degree in the three-phase models owing to trapping of CO in the inert mantle.

The relative surface and mantle populations in the three-phase models F3 and D3 are also displayed in Fig.\,\ref{fig:abundances_1e5_ice_Padovani18Low}. The behavior of the abundances in the reactive surface layer resembles the trends in the time evolution of the gas-phase abundances of relatively volatile molecules after the onset of freeze-out (cf.\,Fig.\,\ref{fig:abundances_1e5_Padovani18Low}). Again, $\rm N_2$ (along with other nitrogen carriers) presents the largest model-dependent differences, reflecting its low reactivity in the ice.

\subsection{Efficiency of radiative cooling}\label{ss:radiativeCooling}

Radiative cooling is inefficient compared to cooling via desorption with the present set of model parameters. Cooling via sublimation is in all of our simulations several orders of magnitude faster than radiative cooling: for 0.1\,$\mu$m grains that are transiently heated to $\sim$70\,K by a CR strike, the rate of energy loss due to radiation is $\approx 1.1 \times 10^{-8} \, \rm erg \, s^{-1}$, which is orders of magnitude lower than the loss rate due to sublimation. \citet{Leger85} predicted that cooling via radiation or via sublimation are comparable in strength if $T_{\rm max} \sim 25 \, \rm K$ (for 0.1\,$\mu$m grains); we have confirmed with our model that such a transition occurs at sub-30\,K values of $T_{\rm max}$. The situation will change drastically if the strong assumptions in the present model regarding the grain heating outlined in Sect.\,\ref{ss:heating}, for example the neglect of the effect of ice mantle in determining $T_{\rm max}$, are relaxed. We will investigate these issues in a future work (Sipil\"a et al., in prep.).

\subsection{Application to the prestellar core L1544}\label{ss:L1544}

\begin{figure*}
	\centering
	\includegraphics[width=1.6\columnwidth]{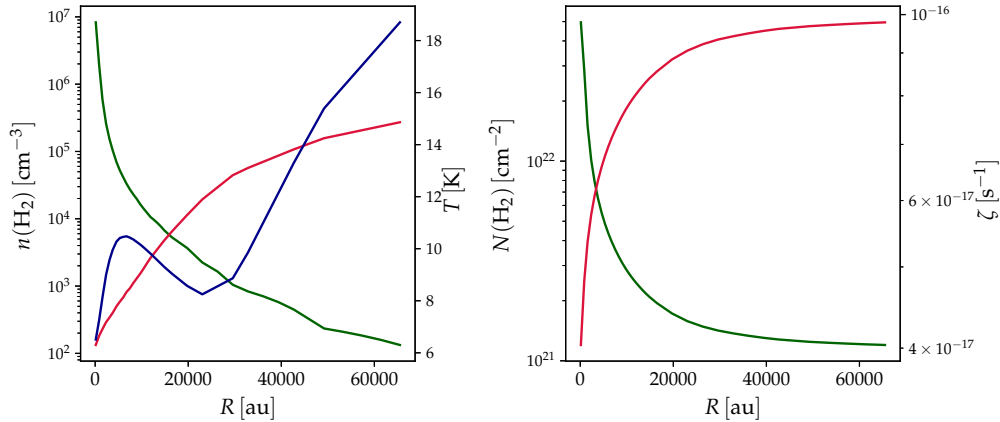}
    \caption{Properties of the L1544 physical model as functions of distance from the core center. {\sl Left}: $\rm H_2$ number density (green), gas temperature (blue), and dust temperature (red). {\sl Right}: $\rm H_2$ column density (green) and CR ionization rate, derived assuming the P18L CR spectrum (red).}
    \label{fig:L1544_properties}
\end{figure*}

It is of interest to investigate the magnitude of the impact of adopting dynamic grain cooling when applied to a starless or prestellar core, where the physical conditions change with location inside the object. We have thus tested our new dynamic CRD scheme in a model of the prestellar core L1544, constructed analogously to the fiducial model described in \citet{Sipila19a}, where a detailed account is given. Briefly, we adopt the source model of \citet{Keto10a}, which gives the density and temperature (gas and dust separately) as a function of radius in the core. We separate the source model into concentric shells, and carry out chemical simulations with our chemical code to derive $r$-dependent chemical abundance profiles that vary with time. The physical structure is kept constant as the chemistry is evolving, i.e., the collapse of the core is not simulated.

An essential difference between the present work and \citet{Sipila19a}, apart from the changes to the CRD scheme, is that here we have derived an $r$-dependent CR ionization rate profile using the formula presented in Appendix~F in \citet{Padovani18}, for the P18L spectrum. However, we multiplied the resulting values by 2 to account for the mirrored CR flux and CRs passing through the center of the cloud \citep{Silsbee18}. The required values of $\tau_{\rm heat}$ were taken from Table~\ref{tab:Rheat} (P18L spectrum) and interpolated to obtain the value corresponding to the $\rm H_2$ column density at each point in the core. Figure~\ref{fig:L1544_properties} shows the physical model parameters as a function of distance from the core center.

\begin{figure*}
	\centering
	\includegraphics[width=1.9\columnwidth]{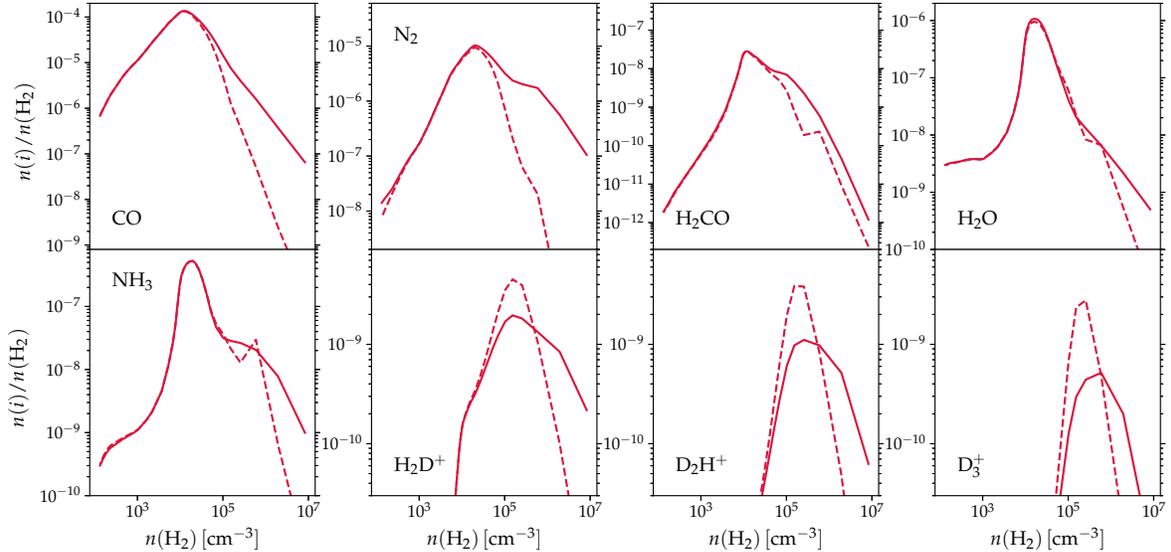}
    \caption{Abundances of selected gas-phase species in the L1544 model as a function of $\rm H_2$ number density at $t = 1.0 \times 10^5 \, \rm yr$. {\sl Solid} lines correspond to model~F2, while {\sl dashed} lines correspond to model~D2.}
    \label{fig:L1544_abundances}
\end{figure*}

We ran the L1544 simulations using the two-phase chemical models F2 and D2. Figure~\ref{fig:L1544_abundances} displays the abundances of the same molecules as in Figs.\,\ref{fig:abundances_1e5_Hasegawa93}~to~\ref{fig:abundances_1e6_Padovani18Low} at time $t = 1.0 \times 10^5 \, \rm yr$ since the beginning of the simulation. The choice of the time step is arbitrary; the abundance plots are provided here for the sake of demonstrating the model-dependent differences, and we do not seek a match to any observational constraints toward L1544. We recover qualitatively similar behavior to the zero-dimensional models; most non-deuterium-bearing molecules experience a drop in abundance when the dynamic CRD scheme is adopted, and the magnitude of the drop again depends on the molecule in question. In line with the zero-dimensional models, the abundances of the deuterated ions are boosted by a factor of a few, or more, when using the dynamic CRD scheme. The abundances of the multiply-deuterated ions are lower in the L1544 model than in the zero-dimensional models. This is because of the decrease of the accretion rates of atomic H and D toward lower temperatures; the associated increase in the gas-phase abundances of these atoms enables the conversion of multiply-deuterated $\rm H_3^+$ back towards $\rm H_2D^+$ which inhibits the formation of other multiply-deuterated ions \citep[see also][]{Sipila15b}. It needs to be noted that the increased abundance of the main deuterated ions does not automatically translate to higher abundances of other deuterium carriers such as deuterated ammonia -- in the case of ammonia, the boost in deuterated ions is offset by the lower gas-phase ammonia abundance, and in fact our model predicts generally somewhat less deuterated ammonia when dynamic CRD is adopted (in two-phase models).

We have tested the effect of limiting the CRD rate of $\rm H_2$, HD, and $\rm D_2$ according to Eq.\,(\ref{kFinal}) in the context of the L1544 model. The high densities and low, sub-10\,K temperatures in the inner core make for an ideal testbed, as the physical conditions are excepted to translate to high abundances of deuterated species, except near the very center where some depletion of deuterated species occurs as well \citep{Redaelli19}. If the desorption rate of HD is limited in the simulations, the abundances of the main deuterated ions drop by several orders of magnitude in the inner core, compared to our fiducial model results, because the temperature is of the order of 7\,K and hence the residence time of atomic and H~and~D is long\footnote{Note that this effect does not occur at 10\,K; cf. our model results in the earlier sections.}. The end result is efficient trapping of HD in the ice and a very strong decrease of deuterium fractionation in the gas phase, which is in complete contradiction with observations toward prestellar cores. The result of this test in combination with the discussion in Sect.\,\ref{ss:H2Cooling} reinforces the notion that more constraints on the role of $\rm H_2$ in grain cooling and its influence on the binding energies of other molecules are sorely needed.

\subsection{Comparison to \citet{Kalvans21}}

\citet{Kalvans21} has recently discussed CRD in the context of a model where the grain cooling time depends on the ice content. They considered grain cooling via the sublimation of CO and $\rm N_2$ ice (in competition with radiative cooling). Their model is very different from ours in many respects; for example they consider, for a given grain size, a range of $T_{\rm max}$ values and derive an average maximum heating temperature that depends on the ice thickness, instead of adopting a single value of $T_{\rm max}$ as we do here. They presented results for an average maximum temperature of 54\,K for 0.1\,$\mu$m grains, leading to grain cooling times that range from a few $\times$ $10^{-3} \, \rm s$ to 10\,s depending on the CO content in the ice.

\begin{figure*}
	\centering
	\includegraphics[width=1.9\columnwidth]{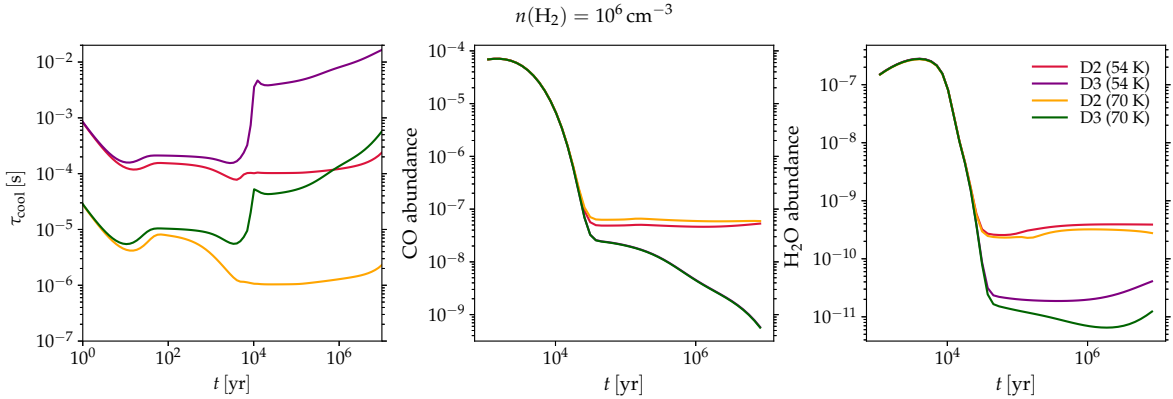}
    \caption{{\sl Left:} Grain cooling time $\tau_{\rm cool}$ as a function of simulation time for models D2~and~D3, assuming $n({\rm H_2}) = 10^6 \, \rm cm^{-3}$ and the P18L CR spectrum, and $T_{\rm max}$ either $54\,\rm K$ (red and purple) or $70\,\rm K$ (orange and green). {\sl Middle:} Gas-phase abundance of CO as a function of time in the four cases. {\sl Right:} As the middle panel, but for $\rm H_2O$.}
    \label{fig:coolingTime_54K}
\end{figure*}

A quantitative comparison of our results to those of \citet{Kalvans21} is challenging owing to the large differences between the model setups, and is beyond the scope of the present paper, but as a test we carried out additional simulations using our high-density physical conditions and adopting $T_{\rm max} = 54\,\rm K$. The resulting grain cooling times are displayed for the P18L CR spectrum in Fig.\,\ref{fig:coolingTime_54K} for our dynamical CRD models D2~and~D3, showing that our models also predict rather long grain cooling times at high densities when $T_{\rm max}$ is decreased from 70\,K, which is a direct consequence of the decreased CRD rates of the cooling molecules. The middle and right-hand panels in Fig.\,\ref{fig:coolingTime_54K} show a sample of the associated effect on chemistry. Evidently the changes induced by decreasing $T_{\rm max}$ from 70\,K to 54\,K are small. The gas-phase water abundance is actually enhanced somewhat for lower $T_{\rm max}$, but we note that this likely a second-order effect, as changing $T_{\rm max}$ influences the grain-surface chemistry due to the longer residence times.

We point out that in these test simulations we employed the grain heating timescales presented in our Table~\ref{tab:Rheat} that are clearly lower than those calculated by \citet{Kalvans21}, which further decreases the compatibility of our respective simulations. Nevertheless it is clear that the maximum transient grain temperature does have an effect on the grain cooling and hence on the overall chemical evolution in interstellar clouds, and should be investigated further.

\section{Conclusions}\label{s:conclusions}

We presented a time-dependent model for the CR-induced desorption of interstellar ices, in which the cooling time of a grain following a CR strike is determined based on the chemical content of the ice at any given time. The model improves on the earlier, widely-adopted CRD model of \citet{Hasegawa93a}, where the cooling time, derived assuming that the ice consists of a homogeneous layer of a generic (CO-like) volatile molecule, is set to a constant value of $10^{-5}\,\rm s$. We varied the time interval between successive CR strikes by considering several CR spectra -- the efficiency of the desorption process is determined by the ratio of the cooling time to the time between successive strikes. We applied the revised CRD scheme to two-phase and three-phase chemical models of gas-grain chemistry in physical conditions corresponding to starless and prestellar cores as well as the envelopes of molecular clouds surrounding the cores. We described the theoretical underpinnings of our work along with the relevant formulae, and noted how the new dynamic CRD scheme can be applied in a typical rate-equation based gas-grain chemical model.

We found that the time-dependent approach to CRD modifies the results in a way that depends on the type of model and on the physical conditions that one considers. In high-density gas, in the case of a two-phase chemical model, grains cool faster as compared to the static CRD scheme, which means that ice molecules have less time to be desorbed, translating to a higher gas-phase depletion factor for the various molecules. However, in a three-phase chemical model in which only the surface ice layer is considered active, the results from the static and time-dependent models are very similar. Switching from the static to the time-dependent CRD scheme has only a very small effect on total (surface+bulk) ice abundances. In physical conditions corresponding to molecular cloud envelopes, clear trends are no longer evident in gas-phase abundances; the differences in the results of static and time-dependent cooling schemes are molecule-dependent.

Our time-dependent CRD scheme is straightforward to apply to any gas-grain chemical model, and we strongly recommend its adoption over the static one that ignores the evolution of the ice content. However, it should be noted that because the grain cooling time is determined by the binding energy of various molecules, our approach is sensitive to uncertainties in the values of the binding energies, and on the type of surface that the molecules are bound to. It is unfortunately difficult to disentangle the effect that varying binding energies have on the cooling time because the binding energy variations have a large impact on the chemistry as a whole.

Much work remains to be done to quantify better the transient heating of dust grains by CR strikes and its effect on ice chemistry. The time interval between successive CR strikes on a grain is affected by the flux of CRs, while the energy deposited into the grain depends on the CR energy spectrum as well as the CR particle composition -- in a realistic scenario, a range of maximum grain temperatures instead of a single value of $T_{\rm max}$ is expected, which has implications for gas-phase and ice chemistry \citep[e.g.,][]{Kalvans19}. The \citet{Hasegawa93a} model for CR-induced desorption is tailored for $T_{\rm max} = 70\, \rm K$ in that the grain cooling time is set to the desorption rate of CO at this temperature. Therefore we expect large differences in simulation results using either our dynamic grain cooling scheme or the constant one of \citet{Hasegawa93a} when a realistic range of $T_{\rm max}$ is considered.

The accumulation of an ice mantle on the grain further modifies the maximum temperature that the grain will reach upon a CR impact in a time-dependent fashion. The problem is highly non-linear, and it is not clear how temporal changes in $T_{\rm max}$ will affect the desorption efficiency and the grain cooling time, and hence the gas-phase and ice abundances, within the context of our new CRD scheme. We will investigate these issues in upcoming works (Silsbee et al., in prep.; Sipil\"a et al., in prep.).

\acknowledgments
We thank the anonymous referees for comprehensive comments that helped to improve the manuscript. The support by the Max Planck Society is gratefully acknowledged.

\appendix

\section{Additional simulation results calculated using the P18H CR spectrum}

\begin{figure*}[b]
	\centering
	\includegraphics[width=0.9\columnwidth]{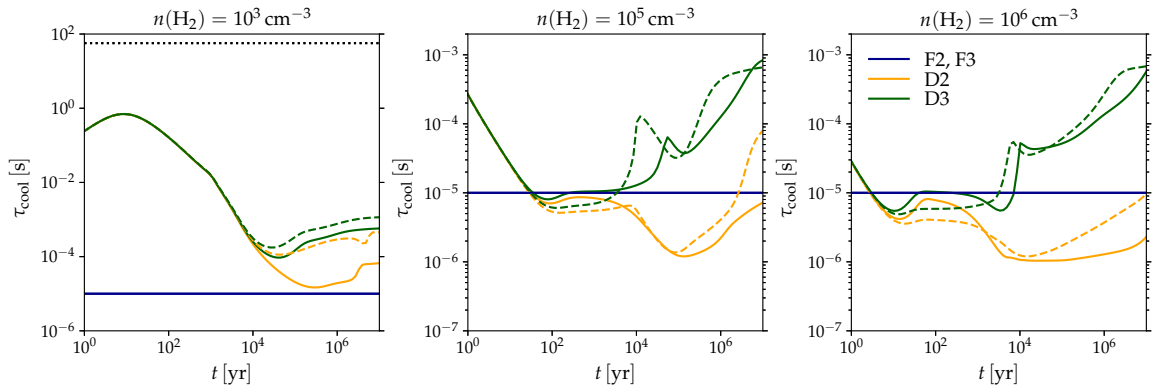}
    \caption{As Fig.\,\ref{fig:coolingTime}, but comparing the grain cooling times calculated using the P18L ({\sl solid} lines) or P18H ({\sl dashed} lines) CR spectrum.}
    \label{fig:coolingTime_P18H}
\end{figure*}

\begin{figure*}
	\centering
	\includegraphics[width=0.9\columnwidth]{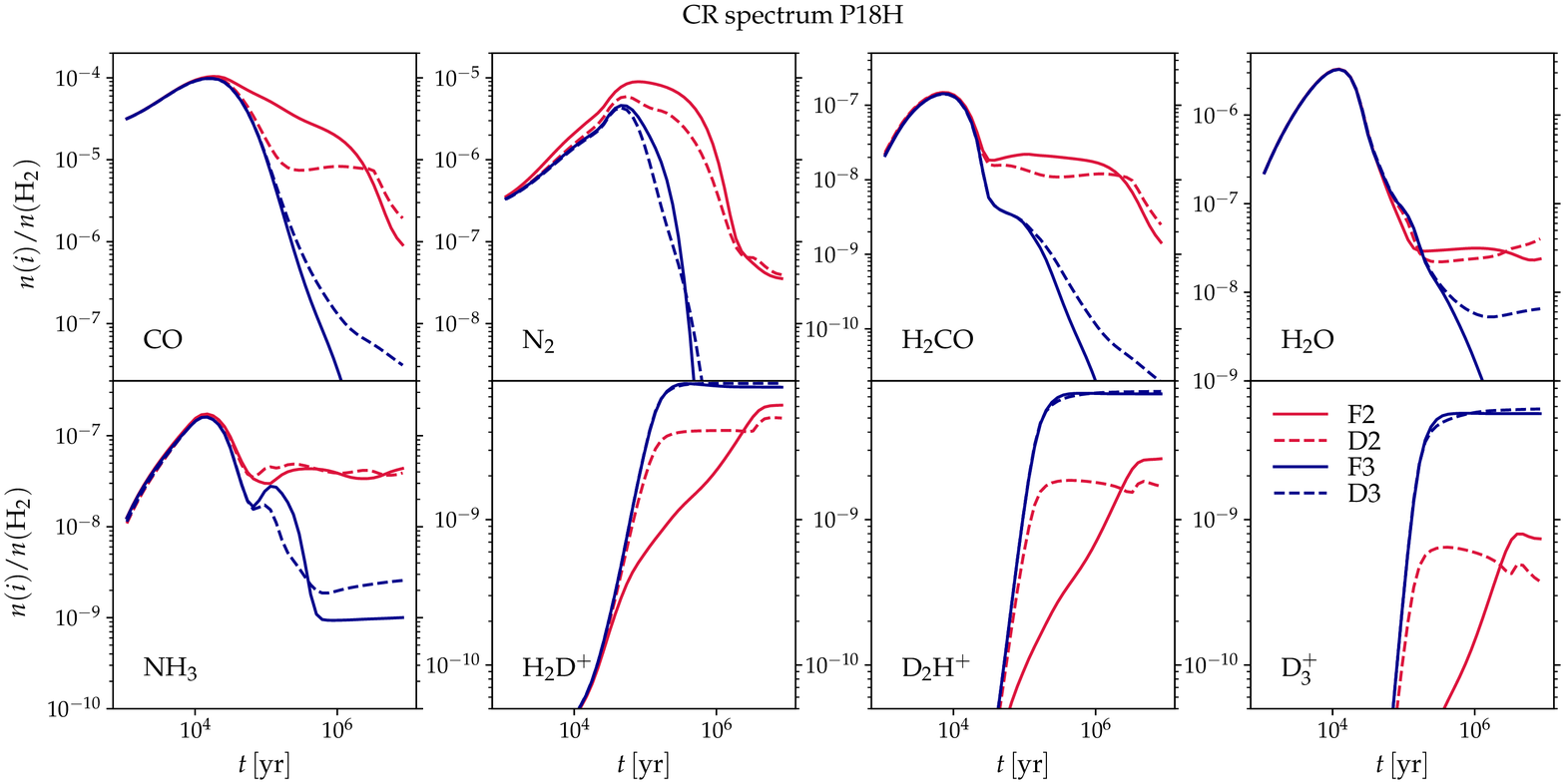}
    \caption{As Fig.\,\ref{fig:abundances_1e5_Padovani18Low}, but calculated using the P18H CR spectrum. Note that the y-axis scaling is different to that in Fig.\,\ref{fig:abundances_1e5_Padovani18Low}.}
    \label{fig:abundances_1e5_Padovani18High}
\end{figure*}

\begin{figure*}
	\centering
	\includegraphics[width=0.9\columnwidth]{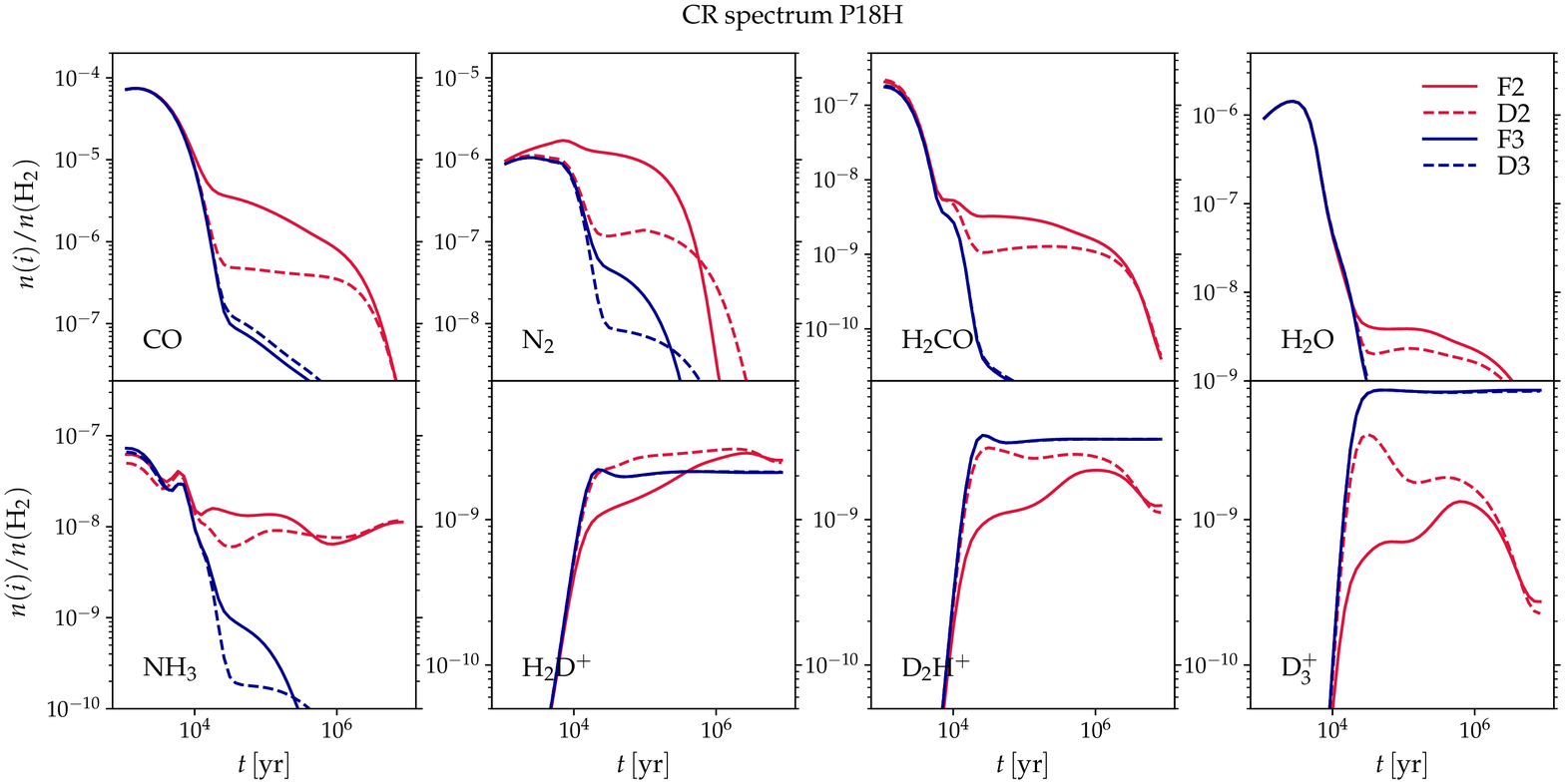}
    \caption{As Fig.\,\ref{fig:abundances_1e6_Padovani18Low}, but calculated using the P18H CR spectrum. Note that the y-axis scaling is different to that in Fig.\,\ref{fig:abundances_1e6_Padovani18Low}.}
    \label{fig:abundances_1e6_Padovani18High}
\end{figure*}

\begin{figure*}
	\centering
	\includegraphics[width=0.95\columnwidth]{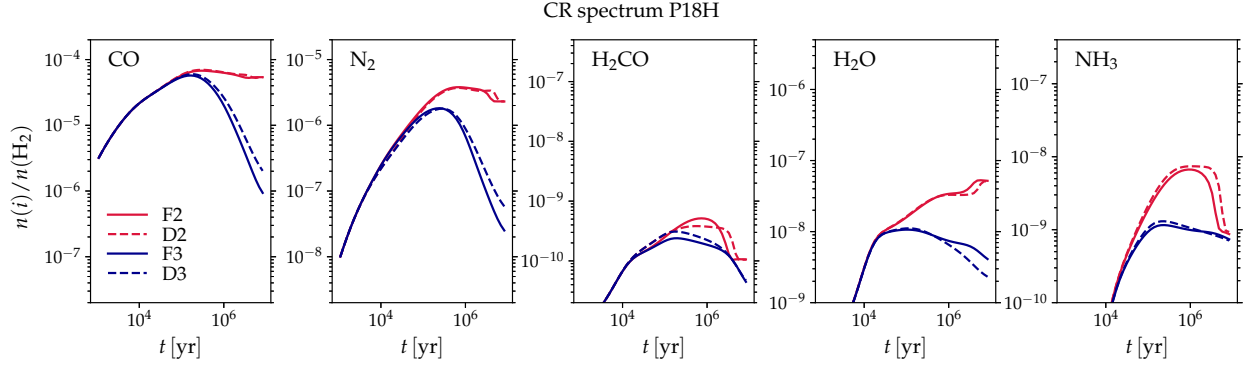}
    \caption{As Fig.\,\ref{fig:abundances_1e3_Padovani18Low}, but calculated using the P18H CR spectrum.}
    \label{fig:abundances_1e3_Padovani18High}
\end{figure*}

Figures~\ref{fig:coolingTime_P18H}~to~\ref{fig:abundances_1e3_Padovani18High} display the effect of the short grain heating interval and increased CR ionization rate due to the high CR flux in model~$\mathscr{H}$ of \citet{Padovani18}. The grain cooling time tends to be longer in these models as compared to the P18L case. The difference is however small enough to not affect the simulation results significantly; our general conclusions presented in main text remain valid for the P18H case as well. We note that the comparison between the P18H and P18L cases is complicated by the fact that the CR ionization rate is higher in the former.

\bibliographystyle{aasjournal}
\bibliography{refs.bib}

\begin{thebibliography}{}
\expandafter\ifx\csname natexlab\endcsname\relax\def\natexlab#1{#1}\fi
\providecommand{\url}[1]{\href{#1}{#1}}
\providecommand{\dodoi}[1]{doi:~\href{http://doi.org/#1}{\nolinkurl{#1}}}
\providecommand{\doeprint}[1]{\href{http://ascl.net/#1}{\nolinkurl{http://ascl.net/#1}}}
\providecommand{\doarXiv}[1]{\href{https://arxiv.org/abs/#1}{\nolinkurl{https://arxiv.org/abs/#1}}}

\bibitem[{{Aguilar} {et~al.}(2015){Aguilar}, {Aisa}, {Alpat}, {Alvino},
  {Ambrosi}, {Andeen}, {Arruda}, {Attig}, {Azzarello}, {Bachlechner}, {Barao},
  {Barrau}, {Barrin}, {Bartoloni}, {Basara}, {Battarbee}, {Battiston}, {Bazo},
  {Becker}, {Behlmann}, {Beischer}, {Berdugo}, {Bertucci}, {Bigongiari},
  {Bindi}, {Bizzaglia}, {Bizzarri}, {Boella}, {de Boer}, {Bollweg},
  {Bonnivard}, {Borgia}, {Borsini}, {Boschini}, {Bourquin}, {Burger}, {Cadoux},
  {Cai}, {Capell}, {Caroff}, {Casaus}, {Cascioli}, {Castellini}, {Cernuda},
  {Cerreta}, {Cervelli}, {Chae}, {Chang}, {Chen}, {Chen}, {Cheng}, {Chen},
  {Cheng}, {Chou}, {Choumilov}, {Choutko}, {Chung}, {Clark}, {Clavero},
  {Coignet}, {Consolandi}, {Contin}, {Corti}, {Gil}, {Coste}, {Creus},
  {Crispoltoni}, {Cui}, {Dai}, {Delgado}, {Della Torre}, {Demirk{\"o}z},
  {Derome}, {Di Falco}, {Di Masso}, {Dimiccoli}, {D{\'\i}az}, {von Doetinchem},
  {Donnini}, {Du}, {Duranti}, {D'Urso}, {Eline}, {Eppling}, {Eronen}, {Fan},
  {Farnesini}, {Feng}, {Fiandrini}, {Fiasson}, {Finch}, {Fisher},
  {Galaktionov}, {Gallucci}, {Garc{\'\i}a}, {Garc{\'\i}a-L{\'o}pez},
  {Gargiulo}, {Gast}, {Gebauer}, {Gervasi}, {Ghelfi}, {Gillard}, {Giovacchini},
  {Goglov}, {Gong}, {Goy}, {Grabski}, {Grandi}, {Graziani}, {Guandalini},
  {Guerri}, {Guo}, {Haas}, {Habiby}, {Haino}, {Han}, {He}, {Heil}, {Hoffman},
  {Hsieh}, {Huang}, {Huh}, {Incagli}, {Ionica}, {Jang}, {Jinchi}, {Kanishev},
  {Kim}, {Kim}, {Kirn}, {Kossakowski}, {Kounina}, {Kounine}, {Koutsenko},
  {Krafczyk}, {La Vacca}, {Laudi}, {Laurenti}, {Lazzizzera}, {Lebedev}, {Lee},
  {Lee}, {Leluc}, {Levi}, {Li}, {Li}, {Li}, {Li}, {Li}, {Li}, {Li}, {Li}, {Li},
  {Lim}, {Lin}, {Lipari}, {Lippert}, {Liu}, {Liu}, {Lolli}, {Lomtadze}, {Lu},
  {Lu}, {Lu}, {Luebelsmeyer}, {Luo}, {Lv}, {Majka}, {Ma{\~n}{\'a}},
  {Mar{\'\i}n}, {Martin}, {Mart{\'\i}nez}, {Masi}, {Maurin}, {Menchaca-Rocha},
  {Meng}, {Mo}, {Morescalchi}, {Mott}, {M{\"u}ller}, {Ni}, {Nikonov},
  {Nozzoli}, {Nunes}, {Obermeier}, {Oliva}, {Orcinha}, {Palmonari},
  {Palomares}, {Paniccia}, {Papi}, {Pauluzzi}, {Pedreschi}, {Pensotti},
  {Pereira}, {Picot-Clemente}, {Pilo}, {Piluso}, {Pizzolotto}, {Plyaskin},
  {Pohl}, {Poireau}, {Postaci}, {Putze}, {Quadrani}, {Qi}, {Qin}, {Qu},
  {R{\"a}ih{\"a}}, {Rancoita}, {Rapin}, {Ricol}, {Rodr{\'\i}guez},
  {Rosier-Lees}, {Rozhkov}, {Rozza}, {Sagdeev}, {Sandweiss}, {Saouter},
  {Sbarra}, {Schael}, {Schmidt}, {von Dratzig}, {Schwering}, {Scolieri}, {Seo},
  {Shan}, {Shan}, {Shi}, {Shi}, {Shi}, {Siedenburg}, {Son}, {Spada},
  {Spinella}, {Sun}, {Sun}, {Tacconi}, {Tang}, {Tang}, {Tang}, {Tao},
  {Tescaro}, {Ting}, {Ting}, {Tomassetti}, {Torsti}, {T{\"u}rko{\v{g}}lu},
  {Urban}, {Vagelli}, {Valente}, {Vannini}, {Valtonen}, {Vaurynovich},
  {Vecchi}, {Velasco}, {Vialle}, {Vitale}, {Vitillo}, {Wang}, {Wang}, {Wang},
  {Wang}, {Wang}, {Wang}, {Weng}, {Whitman}, {Wienkenh{\"o}ver}, {Wu}, {Wu},
  {Xia}, {Xie}, {Xie}, {Xiong}, {Xin}, {Xu}, {Xu}, {Yan}, {Yang}, {Yang}, {Ye},
  {Yi}, {Yu}, {Yu}, {Zeissler}, {Zhang}, {Zhang}, {Zhang}, {Zhang}, {Zheng},
  {Zhuang}, {Zhukov}, {Zichichi}, {Zimmermann}, {Zuccon}, {Zurbach}, \& {AMS
  Collaboration}}]{Aguilar15}
{Aguilar}, M., {Aisa}, D., {Alpat}, B., {et~al.} 2015, \prl, 114, 171103,
  \dodoi{10.1103/PhysRevLett.114.171103}

\bibitem[{{Ave} {et~al.}(2008){Ave}, {Boyle}, {Gahbauer}, {H{\"o}ppner},
  {H{\"o}randel}, {Ichimura}, {M{\"u}ller}, \& {Romero-Wolf}}]{Ave08}
{Ave}, M., {Boyle}, P.~J., {Gahbauer}, F., {et~al.} 2008, \apj, 678, 262,
  \dodoi{10.1086/529424}

\bibitem[{{Balucani} {et~al.}(2015){Balucani}, {Ceccarelli}, \&
  {Taquet}}]{Balucani15}
{Balucani}, N., {Ceccarelli}, C., \& {Taquet}, V. 2015, \mnras, 449, L16,
  \dodoi{10.1093/mnrasl/slv009}

\bibitem[{{Caselli} {et~al.}(1999){Caselli}, {Walmsley}, {Tafalla}, {Dore}, \&
  {Myers}}]{Caselli99}
{Caselli}, P., {Walmsley}, C.~M., {Tafalla}, M., {Dore}, L., \& {Myers}, P.~C.
  1999, \apjl, 523, L165, \dodoi{10.1086/312280}

\bibitem[{{Cernicharo} {et~al.}(2012){Cernicharo}, {Marcelino}, {Roueff},
  {Gerin}, {Jim{\'e}nez-Escobar}, \& {Mu{\~n}oz Caro}}]{Cernicharo12}
{Cernicharo}, J., {Marcelino}, N., {Roueff}, E., {et~al.} 2012, \apjl, 759,
  L43, \dodoi{10.1088/2041-8205/759/2/L43}

\bibitem[{{Crapsi} {et~al.}(2007){Crapsi}, {Caselli}, {Walmsley}, \&
  {Tafalla}}]{Crapsi07}
{Crapsi}, A., {Caselli}, P., {Walmsley}, M.~C., \& {Tafalla}, M. 2007, \aap,
  470, 221, \dodoi{10.1051/0004-6361:20077613}

\bibitem[{{Cummings} {et~al.}(2016){Cummings}, {Stone}, {Heikkila}, {Lal},
  {Webber}, {J{\'o}hannesson}, {Moskalenko}, {Orlando}, \&
  {Porter}}]{Cummings16}
{Cummings}, A.~C., {Stone}, E.~C., {Heikkila}, B.~C., {et~al.} 2016, \apj, 831,
  18, \dodoi{10.3847/0004-637X/831/1/18}

\bibitem[{{Draine}(2011)}]{Draine11}
{Draine}, B.~T. 2011, {Physics of the Interstellar and Intergalactic Medium}
  (Princeton: Princeton Univ. Press)

\bibitem[{{Draine} \& {Bertoldi}(1996)}]{Draine96}
{Draine}, B.~T., \& {Bertoldi}, F. 1996, \apj, 468, 269, \dodoi{10.1086/177689}

\bibitem[{{Furuya} {et~al.}(2015){Furuya}, {Aikawa}, {Hincelin}, {Hassel, G.
  E.}, {Bergin, E. A.}, {Vasyunin, A. I.}, \& {Herbst, E.}}]{Furuya15}
{Furuya}, K., {Aikawa}, Y., {Hincelin}, U., {et~al.} 2015, A\&A, 584, A124,
  \dodoi{10.1051/0004-6361/201527050}

\bibitem[{{Garrod}(2013)}]{Garrod13b}
{Garrod}, R.~T. 2013, \apj, 765, 60, \dodoi{10.1088/0004-637X/765/1/60}

\bibitem[{{Garrod} \& {Pauly}(2011)}]{Garrod11}
{Garrod}, R.~T., \& {Pauly}, T. 2011, \apj, 735, 15,
  \dodoi{10.1088/0004-637X/735/1/15}

\bibitem[{{Garrod} {et~al.}(2007){Garrod}, {Wakelam}, \& {Herbst}}]{Garrod07}
{Garrod}, R.~T., {Wakelam}, V., \& {Herbst}, E. 2007, \aap, 467, 1103,
  \dodoi{10.1051/0004-6361:20066704}

\bibitem[{{Ghesqui{\`e}re} {et~al.}(2018){Ghesqui{\`e}re}, {Ivlev}, {Noble}, \&
  {Theul{\'e}}}]{Ghesquiere18}
{Ghesqui{\`e}re}, P., {Ivlev}, A., {Noble}, J.~A., \& {Theul{\'e}}, P. 2018,
  \aap, 614, A107, \dodoi{10.1051/0004-6361/201732288}

\bibitem[{{Harju} {et~al.}(2008){Harju}, {Juvela}, {Schlemmer}, {Haikala},
  {Lehtinen}, \& {Mattila}}]{Harju08}
{Harju}, J., {Juvela}, M., {Schlemmer}, S., {et~al.} 2008, \aap, 482, 535,
  \dodoi{10.1051/0004-6361:20079259}

\bibitem[{{Hasegawa} \& {Herbst}(1993{\natexlab{a}})}]{Hasegawa93a}
{Hasegawa}, T.~I., \& {Herbst}, E. 1993{\natexlab{a}}, \mnras, 261, 83

\bibitem[{{Hasegawa} \& {Herbst}(1993{\natexlab{b}})}]{Hasegawa93b}
---. 1993{\natexlab{b}}, \mnras, 263, 589

\bibitem[{{Herbst} \& {Cuppen}(2006)}]{Herbst06}
{Herbst}, E., \& {Cuppen}, H.~M. 2006, Proceedings of the National Academy of
  Science, 103, 12257, \dodoi{10.1073/pnas.0601556103}

\bibitem[{{Hollenbach} {et~al.}(2009){Hollenbach}, {Kaufman}, {Bergin}, \&
  {Melnick}}]{Hollenbach09}
{Hollenbach}, D., {Kaufman}, M.~J., {Bergin}, E.~A., \& {Melnick}, G.~J. 2009,
  \apj, 690, 1497, \dodoi{10.1088/0004-637X/690/2/1497}

\bibitem[{{Jim{\'e}nez-Serra} {et~al.}(2016){Jim{\'e}nez-Serra}, {Vasyunin},
  {Caselli}, {Marcelino}, {Billot}, {Viti}, {Testi}, {Vastel}, {Lefloch}, \&
  {Bachiller}}]{Jimenez-Serra16}
{Jim{\'e}nez-Serra}, I., {Vasyunin}, A.~I., {Caselli}, P., {et~al.} 2016,
  \apjl, 830, L6, \dodoi{10.3847/2041-8205/830/1/L6}

\bibitem[{{Kalv{\={a}}ns}(2018)}]{Kalvans18b}
{Kalv{\={a}}ns}, J. 2018, \apjs, 239, 6, \dodoi{10.3847/1538-4365/aae527}

\bibitem[{{Kalv{\={a}}ns}(2021)}]{Kalvans21}
---. 2021, arXiv e-prints, arXiv:2102.02681.
\newblock \doarXiv{2102.02681}

\bibitem[{{Kalv{\={a}}ns} \& {Kalnin}(2019)}]{Kalvans19}
{Kalv{\={a}}ns}, J., \& {Kalnin}, J.~R. 2019, \mnras, 486, 2050,
  \dodoi{10.1093/mnras/stz1010}

\bibitem[{{Kalv{\={a}}ns} \& {Kalnin}(2020)}]{Kalvans20}
---. 2020, \aap, 641, A49, \dodoi{10.1051/0004-6361/202037906}

\bibitem[{{Keto} \& {Caselli}(2010)}]{Keto10a}
{Keto}, E., \& {Caselli}, P. 2010, \mnras, 402, 1625,
  \dodoi{10.1111/j.1365-2966.2009.16033.x}

\bibitem[{{L\'{e}ger} {et~al.}(1985){L\'{e}ger}, {Jura}, \& {Omont}}]{Leger85}
{L\'{e}ger}, A., {Jura}, M., \& {Omont}, A. 1985, \aap, 144, 147

\bibitem[{{Minissale} {et~al.}(2016){Minissale}, {Dulieu}, {Cazaux}, \&
  {Hocuk}}]{Minissale16a}
{Minissale}, M., {Dulieu}, F., {Cazaux}, S., \& {Hocuk}, S. 2016, \aap, 585,
  A24, \dodoi{10.1051/0004-6361/201525981}

\bibitem[{{Najita} {et~al.}(2001){Najita}, {Bergin}, \& {Ullom}}]{Najita01}
{Najita}, J., {Bergin}, E.~A., \& {Ullom}, J.~N. 2001, \apj, 561, 880,
  \dodoi{10.1086/323320}

\bibitem[{{Oba} {et~al.}(2009){Oba}, {Miyauchi}, {Hidaka}, {Chigai},
  {Watanabe}, \& {Kouchi}}]{Oba09}
{Oba}, Y., {Miyauchi}, N., {Hidaka}, H., {et~al.} 2009, \apj, 701, 464,
  \dodoi{10.1088/0004-637X/701/1/464}

\bibitem[{{Padovani} {et~al.}(2018){Padovani}, {Ivlev}, {Galli}, \&
  {Caselli}}]{Padovani18}
{Padovani}, M., {Ivlev}, A.~V., {Galli}, D., \& {Caselli}, P. 2018, \aap, 614,
  A111, \dodoi{10.1051/0004-6361/201732202}

\bibitem[{{Pagani} {et~al.}(2007){Pagani}, {Bacmann}, {Cabrit}, \&
  {Vastel}}]{Pagani07}
{Pagani}, L., {Bacmann}, A., {Cabrit}, S., \& {Vastel}, C. 2007, \aap, 467,
  179, \dodoi{10.1051/0004-6361:20066670}

\bibitem[{{Redaelli} {et~al.}(2019){Redaelli}, {Bizzocchi}, {Caselli},
  {Sipil{\"a}}, {Lattanzi}, {Giuliano}, \& {Spezzano}}]{Redaelli19}
{Redaelli}, E., {Bizzocchi}, L., {Caselli}, P., {et~al.} 2019, \aap, 629, A15,
  \dodoi{10.1051/0004-6361/201935314}

\bibitem[{{Shingledecker} \& {Herbst}(2018)}]{Shingledecker18}
{Shingledecker}, C.~N., \& {Herbst}, E. 2018, Physical Chemistry Chemical
  Physics (Incorporating Faraday Transactions), 20, 5359,
  \dodoi{10.1039/C7CP05901A}

\bibitem[{{Shingledecker} {et~al.}(2019){Shingledecker}, {Vasyunin}, {Herbst},
  \& {Caselli}}]{Shingledecker19}
{Shingledecker}, C.~N., {Vasyunin}, A., {Herbst}, E., \& {Caselli}, P. 2019,
  \apj, 876, 140, \dodoi{10.3847/1538-4357/ab16d5}

\bibitem[{{Silsbee} {et~al.}(2018){Silsbee}, {Ivlev}, {Padovani}, \&
  {Caselli}}]{Silsbee18}
{Silsbee}, K., {Ivlev}, A.~V., {Padovani}, M., \& {Caselli}, P. 2018, \apj,
  863, 188, \dodoi{10.3847/1538-4357/aad3cf}

\bibitem[{{Sipil{\"a}} {et~al.}(2015{\natexlab{a}}){Sipil{\"a}}, {Caselli}, \&
  {Harju}}]{Sipila15a}
{Sipil{\"a}}, O., {Caselli}, P., \& {Harju}, J. 2015{\natexlab{a}}, \aap, 578,
  A55, \dodoi{10.1051/0004-6361/201424364}

\bibitem[{{Sipil{\"a}} {et~al.}(2019{\natexlab{a}}){Sipil{\"a}}, {Caselli}, \&
  {Harju}}]{Sipila19b}
---. 2019{\natexlab{a}}, \aap, 631, A63, \dodoi{10.1051/0004-6361/201936416}

\bibitem[{{Sipil{\"a}} {et~al.}(2019{\natexlab{b}}){Sipil{\"a}}, {Caselli},
  {Redaelli}, {Juvela}, \& {Bizzocchi}}]{Sipila19a}
{Sipil{\"a}}, O., {Caselli}, P., {Redaelli}, E., {Juvela}, M., \& {Bizzocchi},
  L. 2019{\natexlab{b}}, \mnras, 487, 1269, \dodoi{10.1093/mnras/stz1344}

\bibitem[{{Sipil{\"a}} {et~al.}(2015{\natexlab{b}}){Sipil{\"a}}, {Harju},
  {Caselli}, \& {Schlemmer}}]{Sipila15b}
{Sipil{\"a}}, O., {Harju}, J., {Caselli}, P., \& {Schlemmer}, S.
  2015{\natexlab{b}}, \aap, 581, A122, \dodoi{10.1051/0004-6361/201526468}

\bibitem[{{Taquet} {et~al.}(2014){Taquet}, {Charnley}, \&
  {Sipil{\"a}}}]{Taquet14}
{Taquet}, V., {Charnley}, S.~B., \& {Sipil{\"a}}, O. 2014, \apj, 791, 1,
  \dodoi{10.1088/0004-637X/791/1/1}

\bibitem[{{Tielens} {et~al.}(1991){Tielens}, {Tokunaga}, {Geballe}, \&
  {Baas}}]{Tielens91}
{Tielens}, A.~G.~G.~M., {Tokunaga}, A.~T., {Geballe}, T.~R., \& {Baas}, F.
  1991, \apj, 381, 181, \dodoi{10.1086/170640}

\bibitem[{{Tsuge} {et~al.}(2020){Tsuge}, {Hidaka}, {Kouchi}, \&
  {Watanabe}}]{Tsuge20}
{Tsuge}, M., {Hidaka}, H., {Kouchi}, A., \& {Watanabe}, N. 2020, \apj, 900,
  187, \dodoi{10.3847/1538-4357/abab9b}

\bibitem[{{Vastel} {et~al.}(2014){Vastel}, {Ceccarelli}, {Lefloch}, \&
  {Bachiller}}]{Vastel14}
{Vastel}, C., {Ceccarelli}, C., {Lefloch}, B., \& {Bachiller}, R. 2014, \apjl,
  795, L2, \dodoi{10.1088/2041-8205/795/1/L2}

\bibitem[{{Vasyunin} {et~al.}(2017){Vasyunin}, {Caselli}, {Dulieu}, \&
  {Jim{\'e}nez-Serra}}]{Vasyunin17}
{Vasyunin}, A.~I., {Caselli}, P., {Dulieu}, F., \& {Jim{\'e}nez-Serra}, I.
  2017, \apj, 842, 33, \dodoi{10.3847/1538-4357/aa72ec}

\bibitem[{{Vasyunin} \& {Herbst}(2013)}]{Vasyunin13b}
{Vasyunin}, A.~I., \& {Herbst}, E. 2013, \apj, 769, 34,
  \dodoi{10.1088/0004-637X/769/1/34}

\bibitem[{{Vidali} {et~al.}(1991){Vidali}, {Ihm}, {Kim}, \& {Cole}}]{Vidali91}
{Vidali}, G., {Ihm}, G., {Kim}, H.-Y., \& {Cole}, M.~W. 1991, Surface Science
  Reports, 12, 135, \dodoi{10.1016/0167-5729(91)90012-M}

\bibitem[{{Whittet} {et~al.}(1988){Whittet}, {Bode}, {Longmore}, {Adamson},
  {McFadzean}, {Aitken}, \& {Roche}}]{Whittet88}
{Whittet}, D.~C.~B., {Bode}, M.~F., {Longmore}, A.~J., {et~al.} 1988, \mnras,
  233, 321, \dodoi{10.1093/mnras/233.2.321}

\end{thebibliography}

\end{document}